\documentclass{aastex631}

\usepackage{longtable,booktabs}

\begin{document}

\title{The GFCAT: a catalog of ultraviolet variables observed by GALEX with sub-minute resolution}


\correspondingauthor{Chase C. Million}
\email{chase@millionconcepts.com}

\author[0000-0003-2732-3486]{Chase C. Million}
\affiliation{Million Concepts LLC \\
2312 South Preston Street No. 17287 \\
Louisville KY 40217, USA}

\author[0000-0002-7877-3148]{Michael St. Clair}
\affiliation{Million Concepts LLC \\
2312 South Preston Street No. 17287 \\
Louisville KY 40217, USA}

\author[0000-0003-0556-027X]{Scott W. Fleming}
\affiliation{Space Telescope Science Institute \\
3700 San Martin Drive \\
Baltimore MD 21218, USA}

\author[0000-0001-7746-5461]{Luciana Bianchi}
\affiliation{
Department of Physics \& Astronomy \\
Johns Hopkins University \\
3400 North Charles Street \\
Baltimore MD 21218, USA}

\author[0000-0001-5643-8421]{Rachel Osten}
\affiliation{Space Telescope Science Institute \\
3700 San Martin Drive \\
Baltimore MD 21218, USA}
\affiliation{Center for Astrophysical Science \\
Department of Physics \& Astronomy \\
Johns Hopkins University \\
3400 North Charles Street \\
Baltimore MD 21218, USA}

\begin{abstract}

We have performed the first systematic search of the full GALEX data archive for astrophysical variability on timescales of seconds to minutes by rebinning data across the whole mission to 30-second time resolution. The result is the GALEX Flare Catalog (GFCAT) which describes 1426 ultraviolet variable sources, including stellar flares, eclipsing binaries, $\delta$ Scuti and RR Lyrae variables, and Active Galactic Nuclei (AGN). Many of these sources have never previously been identified as variable. We have also assembled a table of observations of ultraviolet flares and accompanying statistics and measurements, including energies, and of candidate eclipsing stars. This effort was enabled by a significantly-enhanced version of the gPhoton software for analyzing time-domain GALEX data; this ``gPhoton2'' package is available to support follow-on efforts. 

\end{abstract}


\section{Introduction} \label{sec:intro}
Many known and theorized astrophysical phenomena can or could only be observed at ``fast'' timescales (seconds to minutes): flares (including `conventional' stellar flares, tidal disruption flares, and shock breakout flares); fast binary systems; fast AGNs; and some astroseismic and starspot activity. Studies of the ``fast'' time-domain radio sky have a particularly long legacy, and the optical and high-energy skies also have decades of time-domain study across multiple facilities. \citep{rau2009exploring, bellm2018zwicky, ricker2014transiting, howard2019evryflare, wevers2018fast} Advancements in materials and data processing capabilities promise to allow instruments like the LSST to detect an even wider scope of transients on the optical and infrared skies. \citep{hambleton2022rubin}

However, time-domain observations are in some sense not repeatable---especially observations of very long-period phenomena and stochastic events like flares---and emerging observational capabilities are unlikely to allow us to take data from the past. New analytic capabilities, however, can allow us to reach into the archive and recover data from the past on scales of a century and down to resolutions of seconds. \citep{grindlay2011opening, mistry2022machine}

Also, due in part to the UV opacity of Earth's atmosphere, we have a much shallower corpus of data on the fast UV sky. This is unfortunate, because the ultraviolet is particularly interesting for the study of stellar phenomena. The relative dimness of cool stars in the UV means that their flaring events have extremely high UV contrast, and many astroseismic and pulsatory events that are poorly understood even within our Sun are principally detectable in transient Lyman emission. \citep{singer2021dorado, lynch2022connecting}

UV transients are also important in the identification of supernova shock breakouts, so-called "orphan" GRBs, and a wide variety of other phenomena. \citep{bayless2022supernova, rhoads1997tell, sagiv2014science} The EUVE, Swift, HST, and GALEX all had or have significant fast capabilities. However, only HST and Swift are still in operation, and only GALEX was designed for wide-field observation. Although many wide-field ultraviolet missions are in development---perhaps most notably ULTRASAT with an anticipated launch in late 2025---the GALEX archival data is the best resource presently available for discovering fast UV events. \citep{ben2022scientific}

\subsection{GALEX}
The Galaxy Evolution Explorer (Martin et al., 2005) was a NASA Small Explorer (SMEX) telescope that surveyed the sky in the ultraviolet over ten years between launch on 28 April 2003 and spacecraft termination on 28 June 2013. GALEX carried two micro-channel plate detectors (MCP) with 1.25 degree fields-of-view (FoV), simultaneously exposed via a dichroic, and capable of observing in either direct imaging or slitless spectroscopic (grism) modes. The detectors observed in two broad ultraviolet (UV) bands centered around 1528 angstroms (Far Ultraviolet or “FUV”) and 2271 angstroms (Near Ultraviolet or “NUV”). The FUV detector failed in May of 2009, but the NUV detector continued to operate until the end of the mission. Individual observations of a particular field (``visits'') lasted between about {1.5} and {26} minutes while the spacecraft was in the shadow of the Earth (an ``eclipse,'' henceforth a ``GALEX eclipse'' to disambiguate from stellar eclipses), once per orbit. GALEX made nearly 30,000 visits before the FUV detector failure, and more than 10,000 after that with only the NUV detector. Many fields were re-visited multiple times.  

The length and breadth of GALEX's observing history has enabled many ``inter-visit'' time-domain analyses on GALEX data (e.g. \citet{gezari2013galex}, \citet{miles2017hazmat}, \citet{szkody2006galex}, \citet{welsh2005galex}, \citet{wheatley2008second}, and \citet{conti2014database}). These studies compare measurements of the same sources across visits, permitting sensitivity to variability on timescales of weeks to years, depending on the revisit rate. Inter-visit analyses typically leverage the mission-produced catalog (``MCAT'') or one of the several derived catalogs of visit-depth photometry, e.g. \citet{seibert2012galaxy}, \citet{bianchi2017revised}, and \citet{olmedo2015deep}.

GALEX is also notable for its high time resolution. Its detectors were ``photon-counting'': unlike many ``integrating'' instruments, they recorded the detector positions of all incident events at a time resolution of at least five-thousands of a second. Moreover, these photon-level data were transmitted fully intact from the spacecraft for reconstruction into visit-depth images on the ground. This permits ``intra-visit'' time-domain analyses with sensitivity to variability on timescales of seconds to minutes.

Intra-visit analyses of GALEX data were substantially simplified by the gPhoton catalog of calibrated GALEX photon events and accompanying software suite \citep{million2016gphoton}. Examples of studies enabled by gPhoton are too numerous to list exhaustively, but include \citet{bianchi2018new}, \citet{de2018searching}, \citet{brasseur2019short}, \citet{rowan2019detections}, \citet{doyle2018stellar}, \citet{boudreaux2017search}, and \citet{jackman2022extending}. Some intra-visit analyses certainly predate gPhoton, including \citet{welsh2005galex}, \citet{wheatley2008second}, \citet{welsh2007detection}, \citet{welsh2011galex}, and \citet{wheatley2012rr}.

Prior to this investigation, however, truly comprehensive intra-visit time-domain studies of GALEX data were impractical. gPhoton had somewhat high runtimes---typically tens of minutes per source---meaning that analyses were limited to small, targeted subsets of data. Following on the effort of \citet{fleming2022new}, in which the function and capabilities of gPhoton were altered in order to enable analyses of flares on nearby GJ65, we have substantially rewritten the gPhoton software. We call the new package ``gPhoton2.'' gPhoton2 features orders-of-magnitude improved run times for creation of both custom lightcurves and calibrated images and movies, as well as the ability to do photometric measurements and lightcurve creation for all sources in a field at once. \citep{stclair_2022_gPhoton2}

We have leveraged the capabilities of gPhoton2 to perform the first comprehensive search of the entire GALEX data archive for intra-visit astrophysical variability. Many of these sources have not been previously identified as variable, and some have been identified as variable with uncertain type identification that this catalog can help refine.

Our interest in the time-resolved morphology of flares in the ultraviolet was the original motivation of this project, whence we arrived at the name---GALEX Flare Catalog or GFCAT. Due to the diversity of intra-visit variability we discovered, we ultimately chose not to restrict the catalog to flares, and about half the final sample of lightcurves do not contain any plausible flaring. However, the catalog does contain 1129 likely flare events, for which we calculated additional quantities: time-resolved FUV / NUV flux ratios (for dual-band observations) and band-encompassed energies (for sources that have a high quality distance measurements).

\section{Data and methods}
\subsection{Initial cut on sample}
Our initial sample of data included all GALEX visits through the end of NASA support for the mission (GALEX eclipse number 46832). We excluded data from the post-NASA ``CAUSE'' phase (``Complete the All-sky UV Survey Extension'')\footnote{http://www.galex.caltech.edu/cause/}, because during CAUSE, new observing modes and relaxed observing safety margins were introduced, which complicates both analysis and interpretation (see, e.g., \citet{olmedo2015deep} and \citet{simons2014ultraviolet}). Also, when there is little time on target, it is difficult to confidently identify astro-physical variability in lightcurves, so we further restricted the search to visits with exposure times of at least 500 seconds conducted in the Medium-depth Imaging Survey (MIS) ``dither'' style of observation. See \citet{million2016gphoton} for a thorough description of the observation modes and surveys. Early in the project, we happened to discover a handful of obviously-variable sources in visits with shorter exposure times; we also included these sources in the final sample. 

\subsection{gPhoton2}
Because the existing gPhoton suite was impractically slow for studying such a large slice of the GALEX corpus, we made a number of architectural and design modifications to gPhoton (henceforth ``gPhoton 1'') to enable large-scale analyses. These optimizations (in large part, leveraging resources that were unavailable, immature, or prohibitively expensive during initial gPhoton 1 development ca. 2012-2015) make \emph{qualitatively} different types of analysis practical. Tasks that would have taken years of compute time can now be performed in hours to days. Its optimizations include: aggressive application of simultaneous programming, including both internal parallelism and highly-configurable, massively distributed processing; the use of fast vectorized operations; locally-cached, conglomerated versions of mission aspect data; moving data processing to local operating environments, thereby avoiding HTTP request overhead; cached just-in-time compilation of Python to C using Numba \citep{lam2015numba}; and heavy use of scratched or shared intermediate products in highly performant storage formats. In aggregate, these modifications improve runtimes by 1-3 orders of magnitude, depending on operation (e.g., first-pass photon calibration on a MIS-like GALEX eclipse typically takes 30-60 seconds rather than 30 minutes).

gPhoton2 also fixes two major bugs in gPhoton. First, although gPhoton had a very robust model of detector dead time, it did not consistently apply it, producing fraction-of-a-second errors in some effective exposure time calculations; this bug often resulted in spurious values for the first and last bins of lightcurves, or for time slices on order of 1-second. Second, gPhoton calculated telemetry flags incorrectly in some edge cases, which caused observation-step-dependent photometric offsets, particularly in FUV. Although these offsets were described in the original gPhoton paper, the fact that they originated in a software bug was only determined during gPhoton2 development. Otherwise, with the exception of a paradigmatic shift from performing photometric operations on measurements individual incident \emph{photons} to vectorized operations on integrated \emph{images}, the underlying algorithms of gPhoton2 are largely unchanged from the original. Aside from improvements created by the flag parsing and dead time fixes, gPhoton2 yields photometric measurements and lightcurves comparable to and consistent with those produced by the ``gAperture'' module of gPhoton 1.

The basic gPhoton2 workflow reduces mission telemetry (``raw6'') files to calibrated photonlists, then scratches the photonlists to disk in Parquet format \citep{vohra2016apache}. Parquet is a columnar data format, which means that subsequent steps of the gPhoton2 workflow can load selected subsets of table fields into memory as needed. It also permits column-wise compression; after extensive benchmarking, we found that the Snappy algorithm \citep{snappy_compression_algorithm} combined with selective dictionary compression provides better tradeoffs between overall data volume and I/O time than `traditional' compression methods like DEFLATE/gzip \citep{gzip_compression}. This scratch step trades a small cost in I/O time for a large savings in working memory, which is important for efficient use of distributed processing resources; also, the photonlists are useful for some types of secondary analysis, and gPhoton2 is capable of directly ingesting them (from local or cloud storage) for subsequent analyses of the same observation to eliminate duplicate processing. Then, gPhoton2 selectively loads fields from these photonlists into memory and uses them to produce full-depth images, along with, optionally, image cubes (``movies'') at a specified time bin size or framerate. As with the photonlists, gPhoton2 is capable of ingesting image/movie products from previous executions to perform extremely fast "photometry-only" runs. The photonlist and image/movie production steps encompass and replace the functionality of the ``gPipeline'' and ``gMap'' modules of gPhoton 1. Aperture photometry is performed on each temporal plane / frame of a ``movie cube.'' gPhoton2's image products include  ``hotspot'' and ``edge'' backplanes that, respectively, mark pixels that contain data covered by the mission-produced hotspot mask or appear within a user-configurable distance of the detector edge. These backplanes have identical dimensions to the image products; when producing lightcurves, gPhoton2 also performs photometry on these backplanes using the same apertures it uses on the image, then automatically propagates these values to hotspot and edge flags on lightcurve bins. This replaces the lightcurve flagging feature of ``gAperture,'' although users may also examine the backplane files to identify per-object and per-bin flags at the pixel level.

\subsection{Source identification and photometric measurement}
gPhoton2's lightcurve production can operate in two modes: one that performs aperture photometry on sources algorithmically identified at runtime, and another that performs photometry on a defined list of sources. We identified most of our sources in the first mode, which analyzes full-depth NUV images with the photutils implementation of the DAOFIND algorithm. \citep{larry_bradley_2019_2533376}, \citep{stetson1987daophot} We varied DAOFIND parameters (``fwhm'' and ``threshold'' in particular) across multiple runs to try to capture all possible sources. We used the second mode to perform targeted searches on known GALEX variables and nearby M Dwarfs (which we expected to have a high rate of detectable flaring and other variability). We compiled this targeted list from \citet{lepine2011all}, \citet{jones2016catalog}, \citet{miles2017hazmat}, \citet{shkolnik2010searching}, \citet{shkolnik2014hazmat}, \citet{welsh2005galex}, \citet{wheatley2008second}, and \citet{gezari2008probing}. We extracted lightcurves from the same source positions in NUV and FUV bands, but performed variability search and screening on NUV and FUV lightcurves separately. We produced lightcurves with 30-second bins, following the recommendation of \citep{million2016gphoton}, and fixed NUV and FUV lightcurves to the same intervals to enable easy cross-comparison.

We produced lightcurves using aperture radii of $9''$, $12.5''$, $17.5''$, $25''$, and $35''$. We provide lightcurves generated using each of these radii in the supplementary data, but except where otherwise noted, we performed all analyses using, and generated all derived quantities from, lightcurves generated from  $17.5''$ radius apertures. Context images provided throughout this paper and included as part of the supplementary data display the $17.5''$ aperture as a solid line and the $25''$ and $35''$ apertures as dashed. 

\subsubsection{Aperture correction}
The estimated aperture correction in both bands for a $17.5''$ aperture is about $7\%$. Although the correction is known to vary across the detector, neither detector-space-dependent adjustments or reliable error estimates are available for the aperture correction. \citep{morrissey2007calibration}  For these reasons, we have chosen to not apply the aperture correction by default to any values reported in GFCAT, except as an intermediate step to computing energies of observed flares (which are sensitive to absolute magnitude). Investigators wishing to attempt to refine the absolute magnitude values provided in GFCAT could apply the $7\%$ aperture correction, use measurements in the $35''$ aperture (which has only an estimated $3\%$ aperture correction and therefore less room for unaccounted error), or attempt to derive their own aperture correction from the multiple measurements provided.

\subsubsection{Background measurement}
We used the $25''$ and $35''$ radii apertures to compute background values, which we then normalized to the area of the $17.5''$ aperture. We provide these background values in the visit-level table, but have not subtracted them from the source fluxes. Catalog users may perform this additional step if they wish. We caution, however, that the GALEX sky background is generally negligible compared to integrated source brightness, and it is very common for background measurements to be dominated by sources near the ``background annulus,'' so background subtraction is unlikely to be useful for most applications. The GALEX sky background may be non-negligible in the presence of cirrus \citep{murthy2014galex} or the extended disks of galaxies, which can be assessed by visual inspection of the images.

\subsection{Variability screening}
Initial photometric extraction generated $\approx420$ million discrete observations of sources (``source-visits'') with accompanying lightcurves. We screened these curves \emph{for} significant variability and \emph{against} common categories of artifacts to ensure, with very high confidence, that every source in the resulting catalog was significantly variable and that its variability had an astrophysical origin. This process ultimately eliminated $\approx99.99995\%$ of source-visits. We provide an outline of our method below, but it is more completely and accurately described \emph{as code} in supplements to this paper.\footnote{https://github.com/millionconcepts/gfcat} \citep{million_2022_gfcat_software}

We used a combination of algorithmic and manual methods to perform screening, starting with the fastest, simplest steps and funneling survivors through more computationally expensive and/or labor-intensive steps. We began with two simple cuts on brightness. To avoid detector non-linearity effects in bright stars that routinely manifest as variability \citep{de2018searching}, we rejected almost all source-visits brighter than 170 counts per second ($\approx14.5$ AB Mag in NUV). We ultimately chose to include a few brighter sources that we are fairly certain exhibit astrophysical variability, but flagged them for possible quality issues. At the lower bound, to reduce the chance of interpreting noise as variability, we rejected source-visits whose lightcurves contained \emph{no} bins brighter than 0.5 cps ($\approx20.8$ AB Mag in NUV), which is roughly GALEX's NUV detection limit within 30-second time bins. 

Our first-pass variability screening on remaining source-visits loosely followed the method outlined in \citet{brasseur2019short}, which treats differences of multiple standard deviations between points in a lightcurve as indicators of ``significant'' variation, with a number of complicating modifications intended to mitigate the rate of false positives. The result corresponds roughly to a detection limit of $\approx3\sigma$ between the dimmest and brightest lightcurve bins. We then used the DBSCAN algorithm to identify and eliminate spatial clusters of sources that vary within the same visit; simultaneous variation of groups of sources is almost certainly due to non-astrophysical effects / artifacts. We performed Anderson-Darling tests \citep{stephens1974} as an additional pruning step to confirm variability, but these had almost total agreement with computationally-cheaper screens based purely on standard deviation.

These algorithmic screening steps produced $\approx20000$ candidate variables. However, there are many forms of non-astrophysical variability in GALEX data, including internal and dichroic reflections, edge effects (e.g. detector regions with poor performance and light leakage from nearby bright stars), detector hotspots and low response regions, transiting artificial satellites, and dither-synchronous pulsations produced by poor response correction or detector nonlinearities. (See Appendix \ref{artifacts_appendix} for more details.) It is often difficult or impossible to identify these artifacts by algorithmic analysis of lightcurves. In order to exclude these artifacts from the catalog, the first author manually reviewed each of these $\approx20000$ source-visits using visual analysis of images, lightcurves, animated movies, cross-catalog comparisons, and ad-hoc numerical analyses. 

We also manually set per-band quality flags for source-visits. These flags indicate that we believe that, although the source-visit contains real astrophysical variability, the corresponding band of that observation is likely contaminated by an artifact or detector issue of some kind (most often a nearby diffuse reflection or hotspot, or a data dropout). We recommend additional caution in analysis of flagged lightcurves. We have, for example, excluded them from the catalog-level analyses of flare properties presented below.

\subsubsection{Crossmatching to Gaia and SIMBAD}
In order to establish prior identifications/classifications for GFCAT sources, we crossmatched them to SIMBAD using a $17.5''$ search radius. We provide the SIMBAD ID for the closest match, along with the distance to that source. We also provide the set of unique SIMBAD ``OTYPE'' values (aka ``object type'') for \emph{all} sources within the search radius and a counter for the total number of sources within that radius. These values should enable users to perform quick, if somewhat naive, assessments of match / identification quality for any specific source in the catalog.

We also crossmatched each GFCAT source to Gaia DR3 using a $17.5''$ search radius. As with the SIMBAD crossmatch, we provide the angular distance to the nearest match source and the total number of sources within the match radius. We provide the parallax and parallax error of the nearest Gaia match, along with the derived distances, for use in subsequent calculation of flare energies. \citep{collaboration2016description, vallenari2022gaia} Because we anticipated that many of the most prominent flare stars would be relatively nearby M dwarfs with high proper motions, we used Gaia-provided proper motions to adjust positions of Gaia sources within 0.05 degrees ($180''$)) of the GFCAT source location to the GALEX observation epoch before crossmatching. However, this had very little effect on the final results.

\begin{table}
	\centering
	\caption{A sample record of 20 unique variable sources identified in GFCAT,
 including a unique GFCAT object identifier, the number of visits included in GFCAT, object type ("OTYPE") as described by SIMBAD for all sources within $17.5''$, and
 the ID given by SIMBAD for the nearest source.}
	\label{tab:object_table}
\begin{tabular}{lcll}
\toprule
GFCAT Object ID &  GFCAT &                      SIMBAD Primary ID &         SIMBAD \\
                &  Visit \# & & OTYPE\\
\midrule
GFCAT J000403.0-044134.3 &        1 &          ATO J001.0125-04.6926 &            Variable* \\
GFCAT J000859.6+233834.9 &        1 &                   ADS   106 AB &                   ** \\
GFCAT J000912.6+425745.1 &        1 &                 TYC 2789-591-1 &              HighPM* \\
GFCAT J000942.5-003347.7 &        1 &          ATO J002.4268-00.5629 &              RRLyrae \\
GFCAT J001049.4-003855.2 &        1 &                             -- &                   -- \\
GFCAT J001141.2-282109.7 &        1 &                             -- &                   -- \\
GFCAT J001611.7-392722.8 &        4 &          CRTS J001611.7-392722 &              RRLyrae \\
GFCAT J001722.4+163035.2 &        1 &                      HD   1309 &              X, Star \\
GFCAT J001823.2+165144.2 &        1 &                             -- &                   -- \\
GFCAT J001838.9-100416.6 &        1 &                  UCAC3 160-834 &                 Star \\
GFCAT J001850.2+161503.0 &        1 &          2XMM J001850.1+161504 &                    X \\
GFCAT J001850.8-722053.7 &        1 & SSTISAGEMA J001851.35-722108.7 & RGB*\_Candidate, Star \\
GFCAT J001934.1-011035.2 &        1 &        2MASS J00193412-0110343 &            Low-Mass* \\
GFCAT J001934.5-230955.0 &        1 &                SIPS J0019-2309 &              HighPM* \\
GFCAT J001949.0-364619.7 &        1 &                             -- &                   -- \\
GFCAT J002018.0-003526.7 &        1 &        2MASS J00201796-0035257 &             delSctV* \\
GFCAT J002031.1-095135.9 &        1 &          ATO J005.1292-09.8597 &                  SB* \\
GFCAT J002111.4-084142.5 &        1 &          1RXS J002112.1-084154 &                    X \\
GFCAT J002144.5+154338.6 &        1 &                             -- &                   -- \\
GFCAT J002242.5-005249.5 &        1 &                             -- &                   -- \\
GFCAT J002504.5-364619.3 &        1 &                      G 267-100 &                   ** \\
GFCAT J002513.2+041156.4 &        1 &                             -- &                   -- \\
GFCAT J002544.1+170453.0 &        1 &                             -- &                   -- \\
GFCAT J002652.1+425002.0 &        1 &                       G 171-59 &              HighPM* \\
GFCAT J002801.1-423812.1 &        1 &                             -- &                   -- \\
GFCAT J002843.2-440024.1 &        3 &                  SV* SON  4791 &              RRLyrae \\
GFCAT J002851.0+410826.0 &        1 &                             -- &                   -- \\
GFCAT J002858.3+013537.2 &        1 &                   HE 0026+0119 &              RRLyrae \\
GFCAT J002918.5-443856.4 &        1 &                             -- &                   -- \\
GFCAT J003028.6-441126.4 &        1 &                             -- &                   -- \\
GFCAT J003039.8-380959.0 &        1 &               UCAC4 260-000506 &            Eruptive* \\
GFCAT J003049.6-420252.9 &        1 &                             -- &                   -- \\
GFCAT J003049.9-715043.0 &        1 & SSTISAGEMA J003050.00-715044.7 &                 Star \\
GFCAT J003220.2-390001.5 &        1 &                             -- &                   -- \\
GFCAT J003257.0+030539.7 &        1 &                             -- &                   -- \\
GFCAT J003336.6-434915.6 &       11 &                    CD-44   135 & Variable*, EmissionG \\
GFCAT J003410.2-120855.5 &        1 &   Gaia DR2 2375643997370225152 &             Star, ** \\
GFCAT J003410.6-440502.6 &        1 &        2MASS J00341047-4405017 &            Eruptive* \\
GFCAT J003442.4-081518.8 &        1 &                             -- &                   -- \\
GFCAT J003515.2-022627.3 &        1 &      GALEX 2673143681802830594 &            blue, QSO \\
\bottomrule
\end{tabular}

\end{table}

\begin{table}
	\centering
	\caption{A sample record of 20 variable source measurements in GFCAT, including GALEX eclipse number right ascension and declination in J2000 decimal degrees, the mean, minimum, and maximum magnitudes during the visit with 1-$\sigma$ errors, and manually set quality flags. The full catalog table available as a supplement to this publication contains far more columns, as described in Section \ref{catalog_description} and completely defined in the supplementary documentation.}
	\label{tab:visit_table}
\begin{tabular}{lrrrrrrrrrr}
\toprule
GALEX & RA    & Dec   & NUV mean & FUV mean & NUV min  & FUV min  & NUV max  & FUV max  & NUV & FUV \\
eclipse & (deg) & (deg) & (AB mag) & (AB mag) & (AB mag) & (AB mag) & (AB mag) & (AB mag) & QA  & QA \\
\midrule
4217 &  80.01 & -49.08 & 19.52$\pm{0.02}$ & 20.30$\pm{0.06}$ & 20.25$\pm{0.21}$ & 21.68$\pm{0.57}$ & 18.41$\pm{0.09}$ & 19.00$\pm{0.20}$ & 0 & 0 \\
4218 & 149.82 &   1.85 & 19.73$\pm{0.03}$ & 21.00$\pm{0.08}$ & 20.24$\pm{0.21}$ & 22.58$\pm{0.77}$ & 19.35$\pm{0.14}$ & 20.06$\pm{0.30}$ & 1 & 0 \\
4231 &  79.29 & -49.24 & 19.27$\pm{0.02}$ & 20.08$\pm{0.06}$ & 20.20$\pm{0.26}$ & 22.47$\pm{0.75}$ & 18.05$\pm{0.08}$ & 18.70$\pm{0.17}$ & 0 & 0 \\
4245 &  79.25 & -48.87 & 16.03$\pm{0.01}$ & 19.91$\pm{0.06}$ & 16.16$\pm{0.03}$ & 20.50$\pm{0.36}$ & 15.76$\pm{0.07}$ & 19.49$\pm{0.24}$ & 0 & 0 \\
4262 & 150.39 &   1.72 & 18.19$\pm{0.01}$ & 21.09$\pm{0.08}$ & 18.65$\pm{0.10}$ & 22.41$\pm{0.73}$ & 17.73$\pm{0.07}$ & 20.12$\pm{0.31}$ & 0 & 0 \\
4278 & 150.39 &   1.72 & 17.68$\pm{0.01}$ & 20.45$\pm{0.06}$ & 18.16$\pm{0.08}$ & 22.03$\pm{0.73}$ & 17.38$\pm{0.06}$ & 19.37$\pm{0.23}$ & 0 & 0 \\
4282 & 150.47 &   2.20 & 19.06$\pm{0.03}$ & 21.07$\pm{0.10}$ & 19.39$\pm{0.14}$ & 22.53$\pm{0.76}$ & 18.57$\pm{0.10}$ & 19.92$\pm{0.29}$ & 0 & 0 \\
4310 & 150.45 &   2.98 & 19.37$\pm{0.02}$ & 20.98$\pm{0.08}$ & 20.21$\pm{0.44}$ & 22.51$\pm{0.76}$ & 18.80$\pm{0.11}$ & 20.05$\pm{0.30}$ & 0 & 0 \\
4322 & 149.97 &   2.78 & 19.72$\pm{0.03}$ & 21.15$\pm{0.08}$ & 20.30$\pm{0.21}$ & 22.58$\pm{0.77}$ & 19.14$\pm{0.13}$ & 20.15$\pm{0.32}$ & 0 & 0 \\
4325 & 149.50 &   2.08 & 19.72$\pm{0.03}$ & 21.19$\pm{0.09}$ & 20.40$\pm{0.22}$ & 26.61$\pm{2.21}$ & 18.26$\pm{0.09}$ & 19.03$\pm{0.20}$ & 0 & 0 \\
4341 & 150.08 &   1.45 & 19.24$\pm{0.03}$ &               -- & 19.70$\pm{0.16}$ &               -- & 18.42$\pm{0.09}$ &               -- & 0 & 0 \\
4355 & 150.47 &   2.20 & 19.02$\pm{0.02}$ & 20.63$\pm{0.07}$ & 19.57$\pm{0.16}$ & 22.49$\pm{0.75}$ & 18.23$\pm{0.14}$ & 19.36$\pm{0.23}$ & 0 & 0 \\
4365 & 150.39 &   1.72 & 17.91$\pm{0.01}$ & 20.89$\pm{0.07}$ & 18.46$\pm{0.10}$ & 22.46$\pm{0.74}$ & 17.52$\pm{0.06}$ & 19.93$\pm{0.29}$ & 0 & 0 \\
4369 & 150.53 &   2.38 & 18.81$\pm{0.02}$ & 19.63$\pm{0.04}$ & 19.76$\pm{0.17}$ & 21.54$\pm{0.54}$ & 17.19$\pm{0.05}$ & 17.79$\pm{0.12}$ & 0 & 0 \\
4406 & 150.54 &   2.12 & 18.95$\pm{0.02}$ & 19.63$\pm{0.05}$ & 20.27$\pm{0.21}$ & 22.39$\pm{0.73}$ & 17.17$\pm{0.05}$ & 17.51$\pm{0.10}$ & 1 & 1 \\
4415 & 150.08 &   1.45 & 19.09$\pm{0.02}$ & 20.59$\pm{0.08}$ & 19.57$\pm{0.15}$ & 22.18$\pm{0.67}$ & 18.20$\pm{0.09}$ & 18.95$\pm{0.19}$ & 0 & 0 \\
4426 & 150.84 &   2.06 & 13.79$\pm{0.00}$ & 18.42$\pm{0.02}$ & 13.87$\pm{0.01}$ & 19.20$\pm{0.21}$ & 13.66$\pm{0.01}$ & 16.83$\pm{0.08}$ & 1 & 0 \\
4469 & 150.42 &   2.13 & 19.51$\pm{0.02}$ & 21.02$\pm{0.07}$ & 20.12$\pm{0.19}$ & 23.22$\pm{0.95}$ & 17.91$\pm{0.07}$ & 18.43$\pm{0.16}$ & 0 & 0 \\
4560 & 162.18 &  58.26 & 19.02$\pm{0.02}$ &               -- & 19.65$\pm{0.16}$ &               -- & 18.55$\pm{0.10}$ &               -- & 1 & 0 \\
4580 & 164.09 &  57.09 & 17.51$\pm{0.01}$ &               -- & 17.83$\pm{0.07}$ &               -- & 17.28$\pm{0.06}$ &               -- & 0 & 0 \\
\bottomrule
\end{tabular}
\end{table}

\begin{figure}
	\includegraphics[width=\columnwidth]{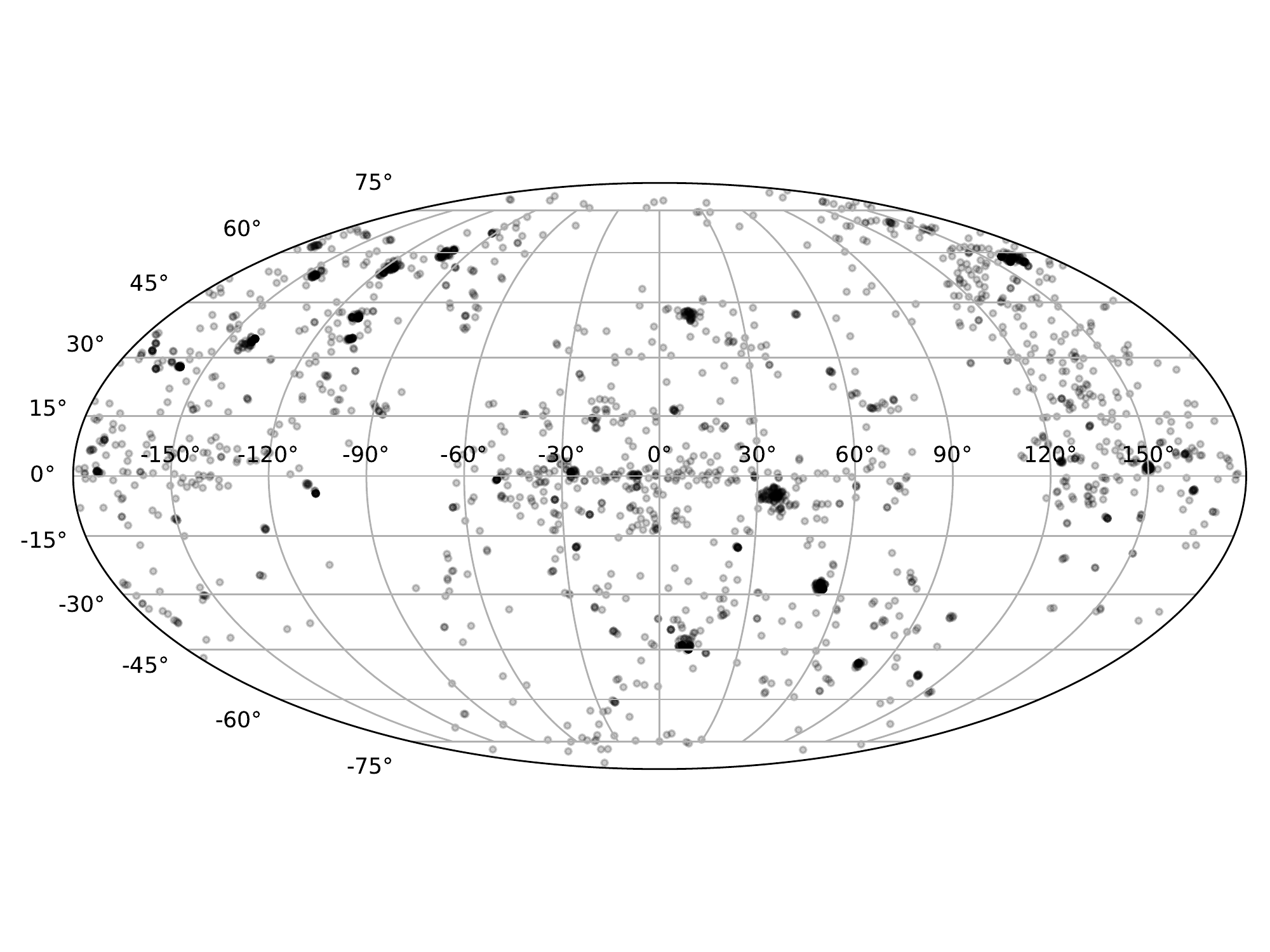}
    \caption{The positions of all sources in GFCAT in right ascension and declination. Darker regions correspond to a higher density of objects. Note that GALEX did not observe the galactic plane or main bodies of the Magellanic clouds until the CAUSE-phase of the mission, which was not included in this effort.}
    \label{fig:total_coverage}
\end{figure}

\begin{figure}
	\includegraphics[width=\columnwidth]{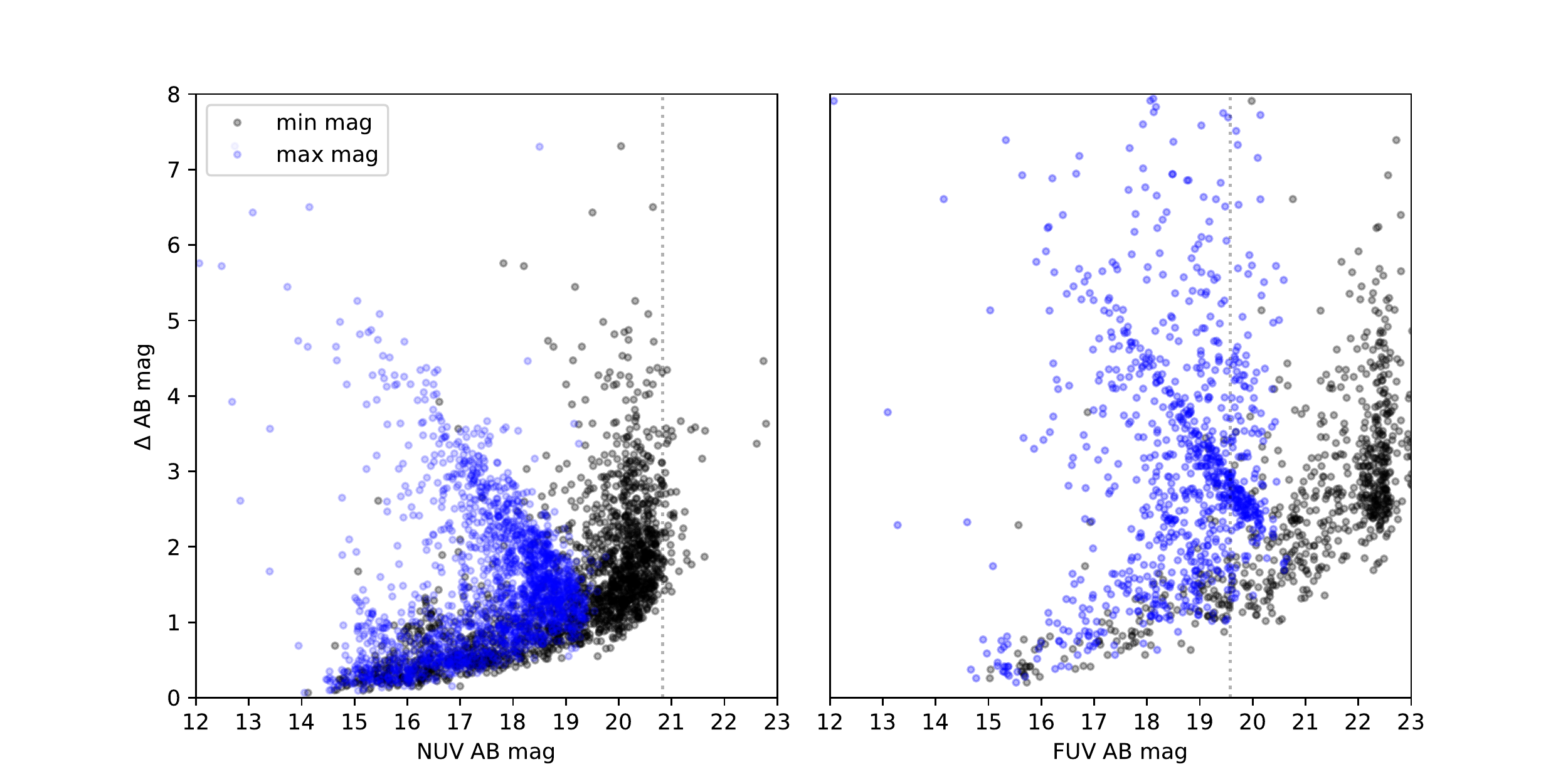}
    \caption{The total magnitude change in each lightcurve ($\Delta$Mag) as a function of the minimum and maximum value in each lightcurve, for both bands. The vertical dotted lines denote $0.5$ counts-per-second equivalent magnitudes. This figure gives a sense of the range of source fluxes contained in GFCAT as well as the detection limits for variability (indicated by the offset between the blue and black distributions as a function of magnitude). The detection limit in NUV is generally consistent with 3-$\sigma$. (See \citet{million2016gphoton} for an exploration of errors as a function of magnitude.) Note that, while we established $0.5$ cps (indicated by the dotted lines) as a soft cutoff for ``detection'' in both bands, stars in our sample are generally dimmer in FUV than NUV. Therefore, many of the variables we detected in NUV are effectively non-detections in FUV.}
    \label{fig:delta_mag_vs_mag}
\end{figure}

\section{Description of the catalog}
\label{catalog_description}
The GFCAT contains 1959 measurements of intra-visit variability across 1426 unique sources. Of these, 588 have not previously been described as variables; the remainder span the gamut of variable stellar types, including flare stars, RR Lyrae, $\delta Scuti$, white dwarf binaries, spectroscopic binaries, and active cool stars. These sources are fairly evenly distributed across the area of the sky GALEX observed in the period of time covered by our sample. Figure \ref{fig:total_coverage} plots the positions of all GFCAT variable sources. Note that this excludes the galactic plane and the cores of the Magellanic clouds. GALEX did not observe these areas until the CAUSE-phase of the mission, which we excluded from this effort. In NUV, lightcurve minimum values range from about 22.5 to 14 AB mag; maximums range from about 19.5 to 12 AB mag. The minimum detected magnitude variation in a lightcurve is 0.06 AB mag (a modest flare on BD+30 1869); the maximum detected variation is 7.3 AB mag (a dramatic flare on G 180-60). Figure \ref{fig:delta_mag_vs_mag} gives a sense of the distribution of variation as well as rough estimates of the sensitivity to variability, which varies as a function of source brightness.

The catalog is composed of four tables. The visit-level table contains statistical information for the 1959 lightcurves / source-visits we identify, with high confidence, as containing astrophysical variability. These statistics include source position; minimum, maximum, and average magnitudes; background estimates; observation times and durations; stellar distances derived from Gaia; stellar identifications (when available) from SIMBAD; and a number of other parameters of interest. A limited subset of the visit-level information is shown in Table \ref{tab:visit_table}. The object-level catalog contains one entry for each of the 1426 unique sources in the GFCAT, along with the source position, stellar identification, and a unique GFCAT ID. A sample of the object-level table is shown in Table \ref{tab:object_table}. The GFCAT also includes separate tables for candidate flare events and candidate stellar eclipse events, which we describe more completely below.

We also provide ``lightcurve files'' for each entry in the visit-level table, which contain lightcurve measurements in both bands (when available) at five separate apertures sizes; and a corresponding QA image which contains a plot of the lightcurve(s), a full-frame image of the observation, and a thumbnail image of the source. These are the same images used during manual QA of the catalog, and provide helpful context for the wide variety of sources in GFCAT. All catalog and lightcurve files, along with complete format descriptions, are provided to MAST as a High Level Science Product via \dataset[10.17909/8d57-1698]{\doi{10.17909/8d57-1698}}. All of the underlying data can be reproduced using gPhoton2 \citep{stclair_2022_gPhoton2} from mission telemetry (``raw6'') files publicly available at the Mikulski Archive for Space Telescopes. We have also made the source code used to generate all results and plots in this paper available, along with a large quantity of supporting software and documentation \citep{million_2022_gfcat_software}.

\subsection{Candidate Flares}
Using the method described in \citet{fleming2022new}, we have calculated additional statistics for the subset of lightcurves manually classified as ``flares.'' This method includes an automated flare detection step, so the lightcurves we manually identified as possibly containing a flare may end up with zero or multiple entries in the flare statistics table, depending on the results of the automated algorithm. Regardless of the results from the automated flare detection algorithm, we retain the ``flare'' classification in the primary catalog even in these cases; like our other type classifications, it is a morphological description and may suggest the presence of flare behavior not captured by the automated detection algorithm.

When both bands are present, we used the temporal extent of the flare as detected in NUV for calculations in both bands. Calculated parameters include estimated ``quiescent'' or instantaneous non-flare flux (INFF), peak flux, duration, integrated fluxes and energies in both bands, and flux ratios (when both bands are available). We have reported fluence for all detections, but because the distance measurement dominates the energy calculation, and can tend quite long for even moderate ($\approx20\%$) errors on parallax, we have declined to calculate energies of flares on any source for which the error on the Gaia parallax is more than 10\%. \citep{luri2018gaia,bailer2015estimating}

In total, we identified 1129 candidate flare events on 883 unique sources across 1011 visits. 540 have dual-band coverage, and we were able to calculate encompassed NUV energies for 949. The lowest-energy flares were measured on the nearby GJ65 binary system (examined in detail in \citet{fleming2022new}), and Wolf 424. This is as expected, because these are the closest active flare stars observed for any significant amount of time by GALEX. The highest energy flare was observed on GUVV J004347.7-445130.7, with an astonishingly high NUV energy of over log$_{10}E=36.2$. The second highest energy flare was observed on TYC 2509-883-1, with an estimated NUV energy of over log$_{10}E=35.4$. We describe both of these flares in more detail in Section \ref{notable_variables}.

Determining, with confidence, that variation in a particular, brief, single-band lightcurves was caused by flaring is a difficult and poorly-constrained problem. It is probably very safe to interpret lightcurves that exhibit the classic Fast Rise Exponential Decay (FRED) morphology as flares, but many likely, known, or suspected flares \emph{do not} exhibit this morphology (see e.g. \citet{howard2022no}). Many types of stars are \emph{capable} of flaring, and stars that are capable of flaring also commonly exhibit other types of short-term variability (e.g. periodic or aperiodic pulsations, other cataclysmic behavior, etc.), so it is difficult to say with certainty---given only a brief GALEX observation---whether any particular bumpy shape in a lightcurve is due to one or more flares or some other type of stellar variation. We have erred towards flagging flare-shaped things as flares, but the list should be approached with caution and skepticism.

\begin{table}
	\centering
	\caption{A sample of data from the table of candidate flare measurements, including source position, distance, estimated durations,
             energies in the bands in which the flare was observed, and a heuristic for whether the detected flare was completely
             captured within the GALEX visit. Energies were only computed for sources with parallaxes measured by Gaia to better than $1\%$ . Any values where the measurements are order-of-magnitude equivalent to their estimated uncertainties can be interpreted as a defacto non-detection, included because it may be useful as a lower limit.}
	\label{tab:flare_table}
\begin{tabular}{lrrrrrrr}

\toprule

       Datetime (ISO) &     RA &    Dec &  distance &  dur. &  NUV energy & FUV energy  & complete \\

                      &  (deg) &  (deg) &   (pc)    &  (s)  & (log10 erg) & (log10 erg) & flare    \\

\midrule

2003-08-03T02:40:09 & 311.57 & -4.93 & 191.54{\raisebox{0.5ex}{\tiny$^{+2.61}_{-2.69}$}} & 1200 & 
33.47{\raisebox{0.5ex}{\tiny$^{+32.23}_{-32.22}$}} & -- & N \\ 

2003-08-03T14:17:46 & 315.94 & -7.38 & 484.88{\raisebox{0.5ex}{\tiny$^{+67.55}_{-93.64}$}} & 300 & 
33.31{\raisebox{0.5ex}{\tiny$^{+32.91}_{-32.91}$}} & -- & Y \\ 

2003-08-03T17:44:22 & 343.35 & -39.79 & 104.51{\raisebox{0.5ex}{\tiny$^{+0.51}_{-0.51}$}} & 450 & 
31.89{\raisebox{0.5ex}{\tiny$^{+30.91}_{-30.90}$}} & -- & N \\ 

2003-08-11T09:35:58 & 259.45 & 59.69 & 129.15{\raisebox{0.5ex}{\tiny$^{+2.03}_{-2.09}$}} & 150 & 
31.38{\raisebox{0.5ex}{\tiny$^{+30.84}_{-30.82}$}} & 30.16{\raisebox{0.5ex}{\tiny$^{+30.53}_{-30.5}$}} & N \\ 

2003-08-17T11:58:32 & 259.20 & 60.04 & 215.61{\raisebox{0.5ex}{\tiny$^{+11.41}_{-12.76}$}} & 210 & 
32.25{\raisebox{0.5ex}{\tiny$^{+31.61}_{-31.58}$}} & -- & Y \\ 

2003-08-18T15:56:17 & 260.31 & 58.10 & 454.26{\raisebox{0.5ex}{\tiny$^{+23.97}_{-26.80}$}} & 420 & 
33.12{\raisebox{0.5ex}{\tiny$^{+32.44}_{-32.41}$}} & 32.62{\raisebox{0.5ex}{\tiny$^{+32.17}_{-32.12}$}} & Y \\ 

2003-08-19T13:19:18 & 258.74 & 58.86 & 470.71{\raisebox{0.5ex}{\tiny$^{+3.12}_{-3.16}$}} & 300 & 
33.66{\raisebox{0.5ex}{\tiny$^{+32.60}_{-32.59}$}} & -- & Y \\ 

2003-08-24T21:35:41 & 340.32 & 12.17 & 286.85{\raisebox{0.5ex}{\tiny$^{+9.24}_{-9.88}$}} & 390 & 
33.10{\raisebox{0.5ex}{\tiny$^{+32.18}_{-32.16}$}} & -- & Y \\ 

2003-08-25T12:23:29 & 350.76 & -0.02 & 506.99{\raisebox{0.5ex}{\tiny$^{+64.96}_{-87.34}$}} & 390 & 
33.52{\raisebox{0.5ex}{\tiny$^{+33.05}_{-33.06}$}} & -- & N \\ 

2003-09-29T14:10:49 & 334.32 & 0.37 & 318.87{\raisebox{0.5ex}{\tiny$^{+2.12}_{-2.15}$}} & 450 & 
32.96{\raisebox{0.5ex}{\tiny$^{+31.90}_{-31.89}$}} & -- & N \\ 

2003-09-30T01:38:10 & 11.24 & -43.40 & 265.09{\raisebox{0.5ex}{\tiny$^{+1.27}_{-1.28}$}} & 510 & 
32.62{\raisebox{0.5ex}{\tiny$^{+31.64}_{-31.63}$}} & 31.55{\raisebox{0.5ex}{\tiny$^{+31.4}_{-31.39}$}} & Y \\ 

2003-10-11T00:42:14 & 21.13 & -33.92 & 25.24{\raisebox{0.5ex}{\tiny$^{+0.01}_{-0.01}$}} & 300 & 
30.52{\raisebox{0.5ex}{\tiny$^{+29.49}_{-29.49}$}} & -- & Y \\ 

2003-10-14T22:25:10 & 30.41 & -8.12 & 88.53{\raisebox{0.5ex}{\tiny$^{+1.40}_{-1.44}$}} & 450 & 
31.71{\raisebox{0.5ex}{\tiny$^{+30.79}_{-30.77}$}} & -- & N \\ 

2003-10-15T04:59:43 & 29.40 & 13.89 & 285.58{\raisebox{0.5ex}{\tiny$^{+7.53}_{-7.95}$}} & 270 & 
32.74{\raisebox{0.5ex}{\tiny$^{+31.87}_{-31.84}$}} & 32.4{\raisebox{0.5ex}{\tiny$^{+31.69}_{-31.66}$}} & Y \\ 

2003-11-25T22:53:12 & 41.63 & -7.35 & 120.16{\raisebox{0.5ex}{\tiny$^{+0.23}_{-0.23}$}} & 480 & 
33.01{\raisebox{0.5ex}{\tiny$^{+31.63}_{-31.62}$}} & 32.48{\raisebox{0.5ex}{\tiny$^{+31.19}_{-31.19}$}} & Y \\ 

2003-12-04T05:47:55 & 52.04 & -28.62 & 281.01{\raisebox{0.5ex}{\tiny$^{+31.97}_{-41.39}$}} & 360 & 
33.20{\raisebox{0.5ex}{\tiny$^{+32.66}_{-32.68}$}} & -- & Y \\ 

2003-12-05T04:48:35 & 52.11 & -28.21 & 510.33{\raisebox{0.5ex}{\tiny$^{+12.07}_{-12.67}$}} & 510 & 
33.76{\raisebox{0.5ex}{\tiny$^{+32.69}_{-32.67}$}} & 33.25{\raisebox{0.5ex}{\tiny$^{+32.4}_{-32.37}$}} & Y \\ 

2003-12-05T04:48:35 & 52.11 & -28.21 & 510.33{\raisebox{0.5ex}{\tiny$^{+12.07}_{-12.67}$}} & 600 & 
33.21{\raisebox{0.5ex}{\tiny$^{+32.37}_{-32.34}$}} & -- & Y \\ 

2003-12-09T18:56:26 & 36.28 & -4.48 & 187.79{\raisebox{0.5ex}{\tiny$^{+3.62}_{-3.76}$}} & 270 & 
32.12{\raisebox{0.5ex}{\tiny$^{+31.33}_{-31.30}$}} & -- & Y \\ 

2004-01-11T17:52:55 & 119.86 & 46.64 & 177.09{\raisebox{0.5ex}{\tiny$^{+3.95}_{-4.13}$}} & 660 & 
32.87{\raisebox{0.5ex}{\tiny$^{+31.83}_{-31.82}$}} & 32.37{\raisebox{0.5ex}{\tiny$^{+31.57}_{-31.55}$}} & N \\ 

\bottomrule

\end{tabular}

\end{table}

\begin{figure}
	\includegraphics[width=\columnwidth]{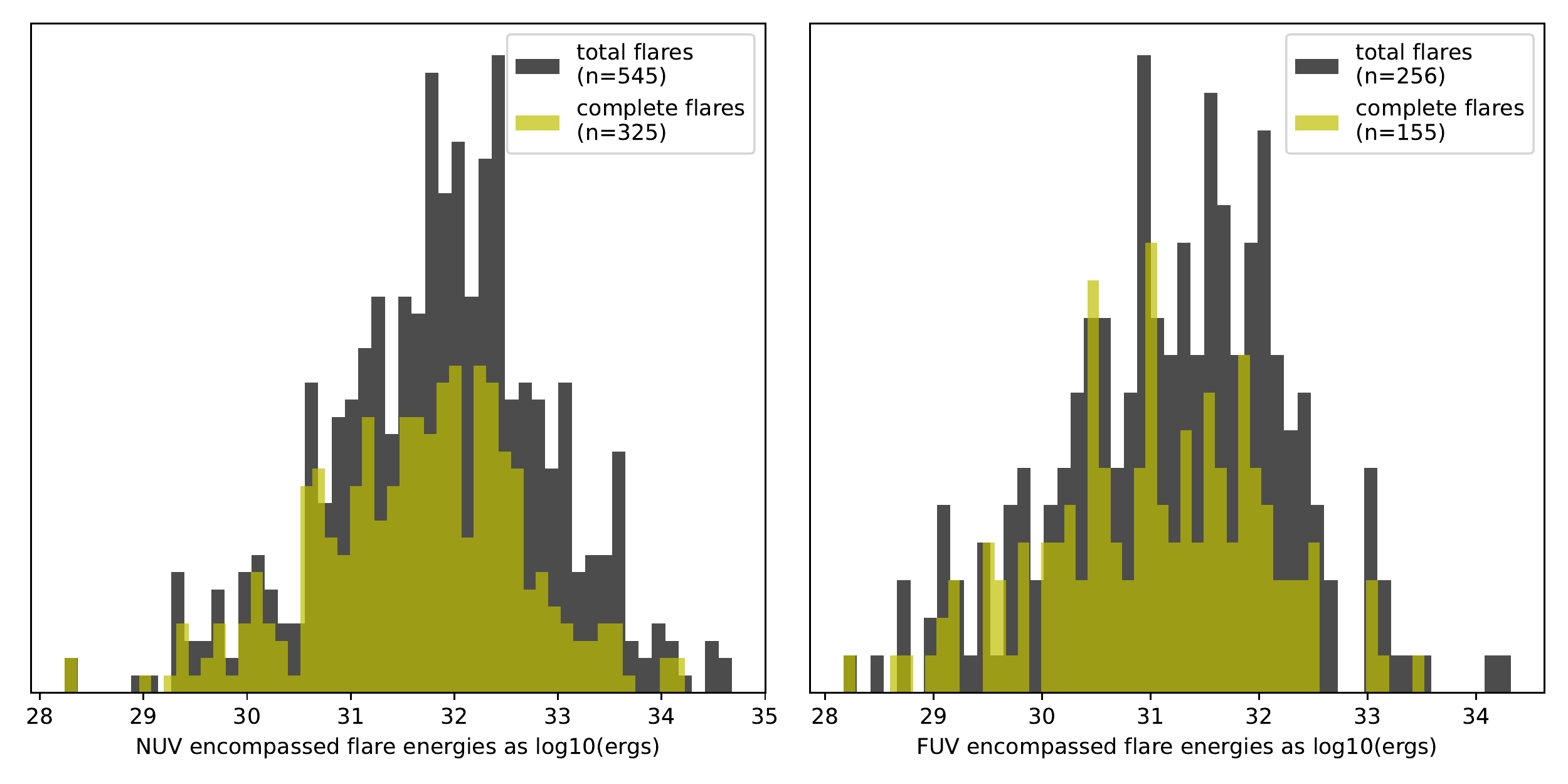}
    \caption{The distribution of measured energies of candidate flares in both bands. ``Complete'' flares are those that have periods of flux consistent with \emph{presumed} quiescence both before and after the identified temporal extent of the candidate flare. Energy measurements for flares that are not ``complete'' should be understood only as lower bounds.}
    \label{fig:example_figure}
\end{figure}

\subsection{Candidate Stellar Eclipses}
We have also manually tagged a subset of visits as ``eclipsing'' (or a ``stellar eclipse,'' which should not be confused with a ``GALEX eclipse''). These lightcurves exhibit very sudden and substantial changes in brightness followed and preceded by relatively flat stretches of quiescence. Although some of the sources we have tagged as eclipses are \emph{known} to be eclipsing or binary stars (per the SIMBAD crossmatch) others are not. These might represent new discoveries of eclipsing binary stars. However, they might just as well be other types of dramatic variable (e.g. $\delta$ Scuti or AGN), many of which can take on a wide variety of temporal morphologies; they might also be brightness changes due to star spots or unusual flaring behavior. So as with our categorization of ``flares'' above, our categorization of ``eclipsing'' sources should be taken as suggestive but not definitive.

Many of these stars suddenly ``vanish'' or ``appear'' during the observation, suggesting that their brightness has dropped below the detection limit. Therefore, the minimum brightness values for these lightcurves are often only upper limits. Investigators can assess this by comparing the measured source flux to the estimated background flux or visual inspection of movie frames.

\begin{table}
	\centering
	\caption{Example of candidate GALEX visits that may contain stellar eclipses.}
	\label{tab:eclipse_table}
\begin{tabular}{lrrrrr}
\toprule
       Datetime (ISO) &     RA &    Dec &          SIMBAD OTYPE &  NUV min. &  NUV max.  \\
            & (deg) & (deg) &       & (AB mag) & (AB mag) \\
\midrule
2006-05-12T08:43:41 & 218.60 &  34.31 &                  Star &        14.76$\pm{0.02}$ &        14.62$\pm{0.02}$ \\
2011-04-07T10:14:58 & 200.62 &   1.01 &            WhiteDwarf &        20.54$\pm{0.55}$ &        18.68$\pm{0.10}$ \\
2008-04-21T04:30:54 & 193.90 &  27.06 & Galaxy, HighPM*, Star &        17.61$\pm{0.06}$ &        15.84$\pm{0.06}$ \\
2009-01-20T02:26:15 & 127.89 &  32.46 &                  Star &        14.83$\pm{0.06}$ &        14.59$\pm{0.02}$ \\
2008-08-19T03:35:59 & 320.05 &  -0.56 &                       &        19.95$\pm{0.18}$ &        17.54$\pm{0.06}$ \\
2004-07-16T15:32:39 & 338.88 &  14.48 &                    ** &        20.17$\pm{0.20}$ &        18.08$\pm{0.08}$ \\
2009-07-26T09:44:45 & 320.64 &  -6.31 &            WhiteDwarf &        20.05$\pm{0.19}$ &        18.78$\pm{0.11}$ \\
2009-06-18T13:00:23 & 207.74 &  26.56 &            Cepheid, X &        19.24$\pm{0.13}$ &        18.02$\pm{0.08}$ \\
2005-01-17T01:03:44 & 160.99 &  58.13 &       CataclyV*, Star &        19.82$\pm{0.17}$ &        17.66$\pm{0.07}$ \\
2006-01-13T03:56:00 & 160.99 &  58.13 &       CataclyV*, Star &        19.88$\pm{0.18}$ &        17.77$\pm{0.07}$ \\
2004-07-24T14:07:05 & 338.88 &  14.48 &                    ** &        20.11$\pm{0.19}$ &        18.03$\pm{0.08}$ \\
2011-02-01T06:43:59 & 154.58 &   8.30 &                       &        20.09$\pm{0.19}$ &        18.87$\pm{0.11}$ \\
2005-03-02T01:59:43 & 160.99 &  58.13 &       CataclyV*, Star &        20.15$\pm{0.20}$ &        17.27$\pm{0.06}$ \\
2005-07-27T18:47:31 & 261.00 &  60.75 &  WhiteDwarf\_Candidate &       20.26$\pm{0.21}$ &        18.94$\pm{0.12}$ \\
2010-07-02T10:37:48 & 292.67 &  40.87 &                  Star &        17.02$\pm{0.06}$ &        15.89$\pm{0.03}$ \\
2006-07-11T03:28:55 & 254.28 &  -4.35 &             Eruptive* &        18.65$\pm{0.11}$ &        16.50$\pm{0.09}$ \\
2011-10-07T13:13:23 & 346.26 &  17.41 &                       &        19.71$\pm{0.16}$ &        18.23$\pm{0.09}$ \\
2010-02-07T14:14:37 & 158.89 &   5.87 &             CataclyV* &        19.86$\pm{0.17}$ &        18.70$\pm{0.11}$ \\
2007-02-22T05:17:04 & 137.05 &   6.07 &             Variable* &        19.98$\pm{0.18}$ &        17.01$\pm{0.05}$ \\
2011-12-20T05:42:03 &  77.16 & -24.20 &                PulsV* &        17.64$\pm{0.07}$ &        16.67$\pm{0.04}$ \\
\bottomrule
\end{tabular}
\end{table}

\section{Variables of note}
\label{notable_variables}
There are more than $1400$ unique variable sources identified in GFCAT. Analyzing them all carefully is reserved for follow-on work. Here are a few brief descriptions of interesting variables that give a taste for the wide range of phenomena represented in GFCAT.

\subsection{Superflare on GUVV J004347.7-445130.7 (GFCAT J004347.8-445132.5)}
\begin{figure}
	\includegraphics[width=0.8\columnwidth,keepaspectratio]{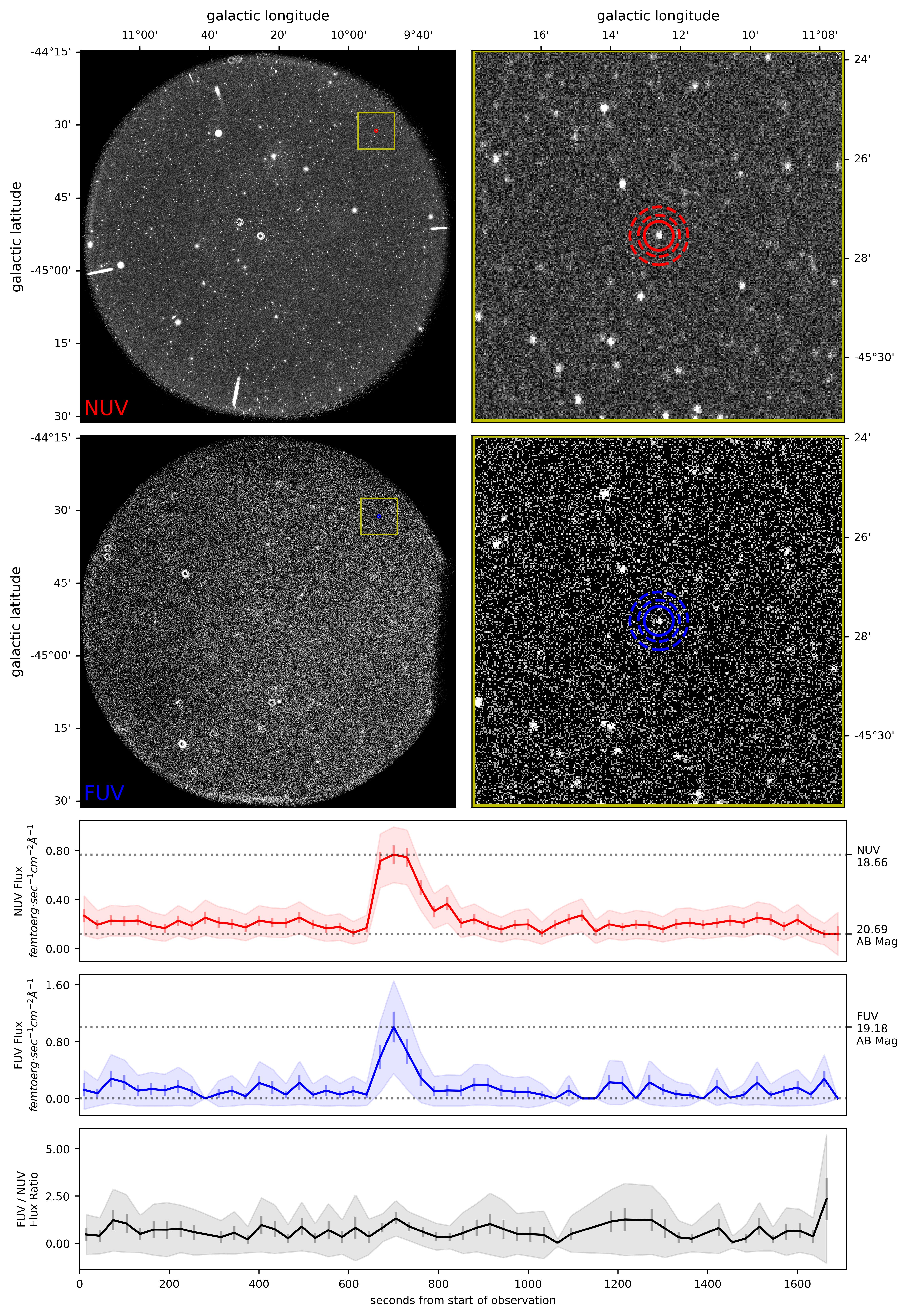}
    \caption{This flare on GUVV J004347.7-445130.7, observed once in GALEX eclipse 7500, is the highest energy flare known to have been observed by GALEX. The top left image is the GALEX NUV full-frame, with the subframe indicated appear in the top right. The images below it are the corresponding FUV full-frame and subframe. The solid circles in the subframe images indicates the $17.5''$ aperture; the dashed lines correspond to $25''$ and $35''$ radii apertures, used for background calculations. The first plot (red line) is the NUV light curve with 30-second bins; the axis on the left is in flux units and the minimum and maximum AB mag are indicated on the right. The second plot (blue line) is the FUV lightcurve with 30-second bins. The bottom plot (black line) is the FUV / NUV flux ratio (unitless).}
    \label{fig:gfcat_07500_0205}
\end{figure}

A flare on GUVV J004347.7-445130.7 (GFCAT J004347.8-445132.5), shown in Figure \ref{fig:gfcat_07500_0205}, is the highest energy flare known to have been observed by GALEX, with an estimated NUV energy of at least
log$_{10}E=36.2${\raisebox{0.5ex}{\tiny$^{+35.4}_{-35.3}$}}. Using the conversion factor of $p_{bol}=0.132$ calculated by \citet{brasseur2019short} for conversion from GALEX NUV bandpass energies to bolometric energies, this suggests a bolometric flare energy of $\approx$log$_{10}E=37.1$. Note that the quiescent flux in both bands is consistent with a non-detection, so this calculation probably provides only a lower bound on energy. This event was identified in the GALEX Ultraviolet Variable (GUVV) catalog \citep{welsh2005galex}, but was not confirmed as a flare, and an energy calculation was not possible because no estimate of distance was available at the time. Gaia puts it at about $2824${\raisebox{0.5ex}{\tiny$^{+193}_{-170}$}} parsecs with about 7\% error on the parallax measurement.

The source was identified as a $\delta$ Scuti variable by \citep{kinman2014identification}. GALEX observed this source 33 times (in MIS mode and not right on the detector edge); 19 of these observations appear in GFCAT, which is to say that they contain significant variability. The lightcurve from this visit---and \emph{only} this visit---has the classical FRED-like shape associated with flares. The other visits show relatively softer trends more consistent with $\delta$ Scuti pulsations. Although they are rare, many observations exist of flares on $\delta$ Scuti stars, and these flares tend to have very high energies. \citep{guzik2021highlights, balona2019evidence, balona2015flare, balona2012kepler} The flux ratio of $\approx0.6$ at flare peak is unusual (but not unheard of) among other flares in our sample, which tend to have a flux ratio around 1.

\subsection{Superflare on previously unknown flare star TYC 2509-883-1 (GFCAT GFCAT J100642.9+365649.0)}
\begin{figure}
	\includegraphics[width=\columnwidth]{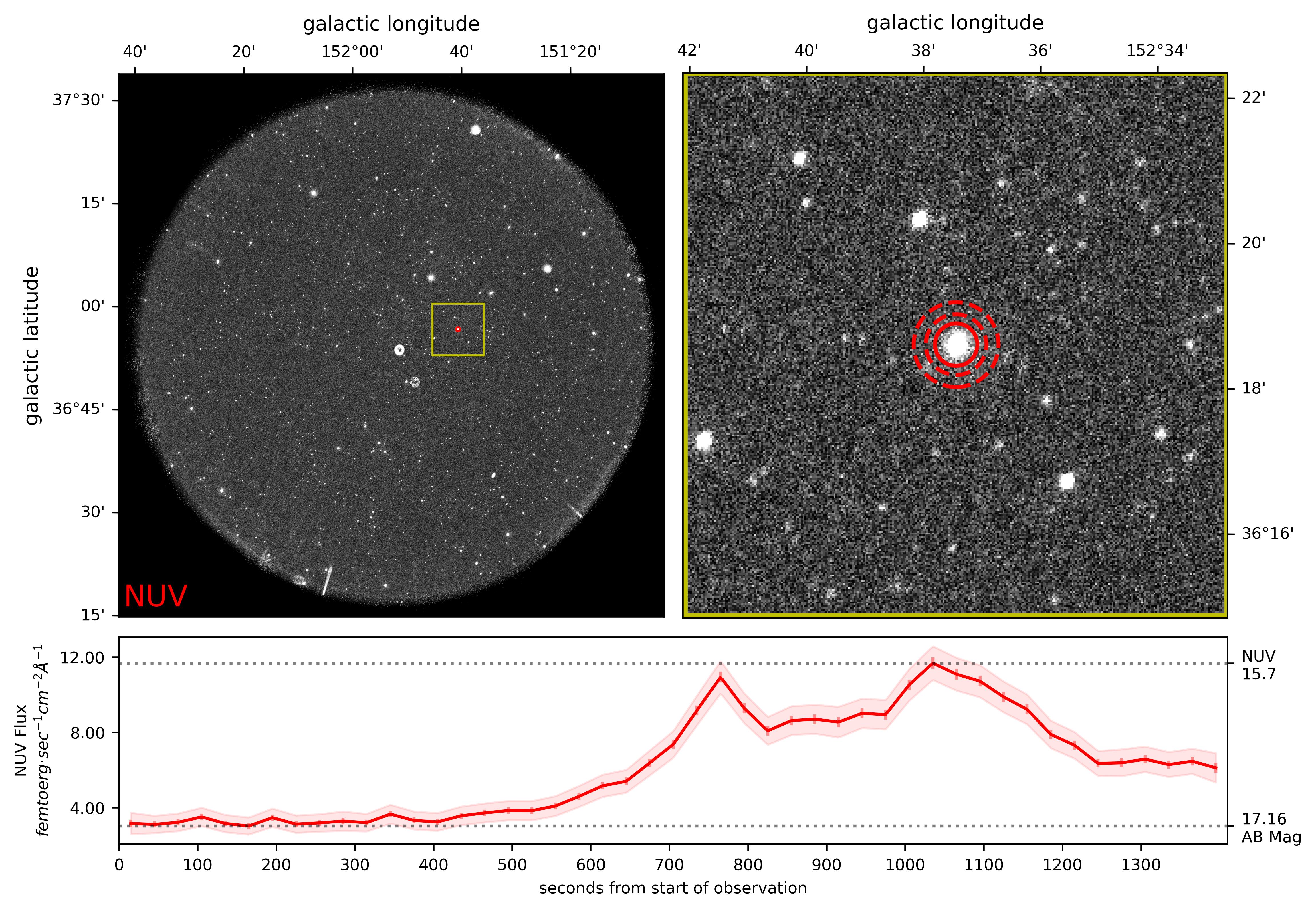}
    \caption{This flare on TYC 2509-883-1, observed once in GALEX eclipse 37332, is the second highest energy flare known to have been observed by GALEX. This star is not previously known to be variable. Refer to the caption of Figure \ref{fig:gfcat_07500_0205} for a description of the plot layout.}
    \label{fig:gfcat_37332_1869}
\end{figure}

Figure \ref{fig:gfcat_37332_1869} shows the second highest energy flare known to have been observed by GALEX, with energy in the NUV band of 
log$_{10}E=35.4${\raisebox{0.5ex}{\tiny$^{+34.2}_{-34.2}$}}. The host star, TYC 2509-883-1 (GFCAT J100642.9+365649.0), is about 900 parsecs distant and not previously known to be variable. The lightcurve of this flare shares an unusual double-peak morphology, also seen in the ``megaflare'' GALEX observed on on GJ 3685A (previously described in \citet{robinson2005galex}); this may suggest a common physical mechanism for these rare, extremely high energy events.

\subsection{Likely stellar eclipses on previously unknown variable GFCAT J223049.4-331941.0}
\begin{figure}
	\includegraphics[width=0.8\columnwidth,keepaspectratio]{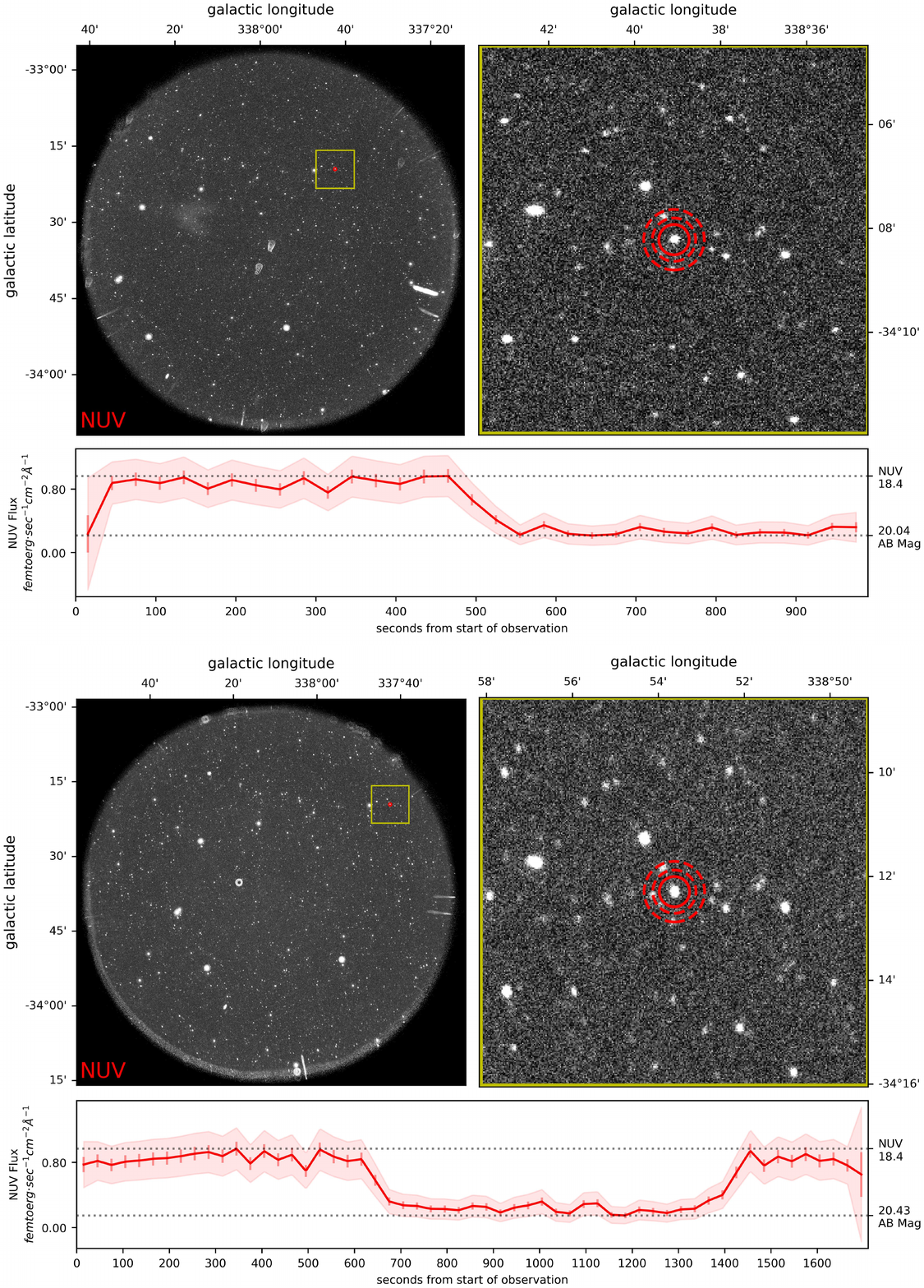}
    \caption{An object observed twice by GALEX, about ten years apart, in GALEX eclipses 2609 (on the top) and 44383 (on the bottom). It has not been previously described, as far as we know, and demonstrates clear stellar eclipse-like behavior. The second visit to the object captured a full eclipse with a duration of approximately 14.5 minutes, including 1-1.5 minute long entry and exit phases, with a depth of at least two AB mag (the lower limit being consistent with a non-detection). Refer to the capture of Figure \ref{fig:gfcat_07500_0205} for a description of the plot layouts.}
    \label{fig:GFCAT_J0223049.4-331941.0}
\end{figure}

The object GFCAT J223049.4-331941.0 shown in Figure \ref{fig:GFCAT_J0223049.4-331941.0} was observed twice, about ten years apart, in GALEX eclipses 2609 and 44383, both times only in NUV. This object demonstrates clear stellar eclipse-like behavior, and as far as we know, it has not previously been described. The second visit to the object captured a full stellar eclipse with a duration of approximately 14.5 minutes, including 1-1.5 minute long entry and exit phases, with a depth of at least 2 AB mag (the lower limit was probably below detection threshold).

\subsection{FUV-only variations on TYC 4988-137-1 (GFCAT J143012.3-021249.2)}
\begin{figure}
	\includegraphics[width=0.8\columnwidth,keepaspectratio]{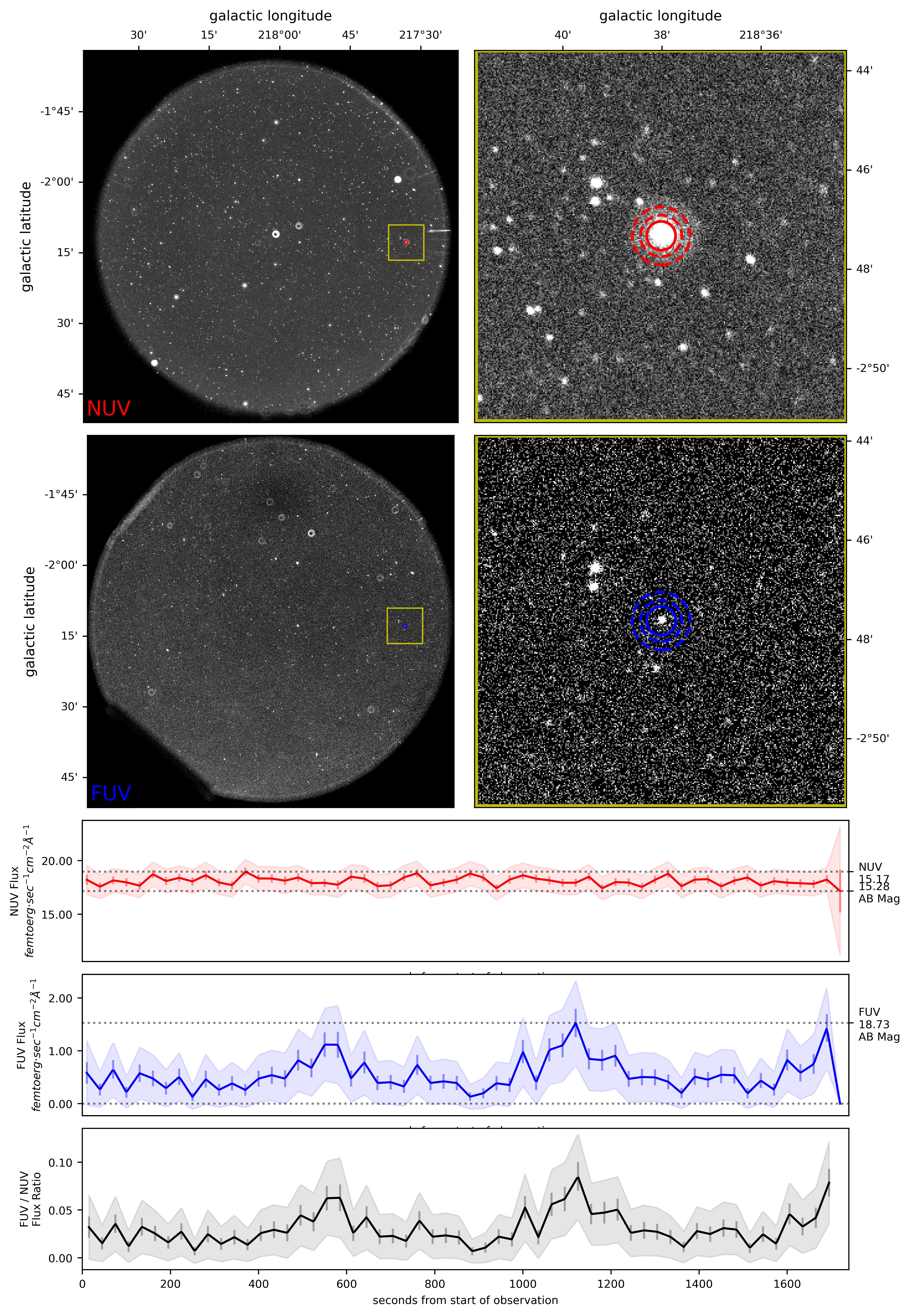}
    \caption{Typically, throughout GFCAT, when a source is detected in both bands, it also exhibits variability in both bands; GFCAT J143012.3-021249.2, with one dual-band observation in GALEX eclipse 26691 shown in Figure \ref{fig:gfcat_26691_0993} is a rare case in which the FUV band shows significant variability and the NUV band---with a very significant detection--shows none. Refer to the capture of Figure \ref{fig:gfcat_07500_0205} for a description of the plot layout.}
    \label{fig:gfcat_26691_0993}
\end{figure}

Typically, throughout GFCAT, sources detected in both bands also vary in both bands. GFCAT J143012.3-021249.2, observed once in both NUV and FUV during GALEX eclipse 26691 and shown in Figure \ref{fig:gfcat_26691_0993}, is a rare case in which the FUV band shows significant variability and the NUV band---despite a very significant detection--shows none. This high proper motion star has been the target of Radial Velocity Experiment (RAVE) studies (e.g. \citet{zwitter2008radial}), but has not previously been described as variable. The FUV source position appears in proximity to one of the ``shadow'' regions of the detector described in the Appendix \ref{fuv_shadow}, which is cause for caution, but it is on the limb of that region and the lightcurve morphology is unlike the ``ramp up`` behavior typically associated with these artifacts.

\section{Conclusions}
GFCAT provides rich opportunities for future investigation. Its sample of 1959 lightcurves from 1426 variable sources encompasses, at minimum, many examples of previously unknown variable stars, newly discovered eclipsing binary systems, and the highest-energy flare ever observed in the ultraviolet. There is a great deal of room for investigating these sources further with the GALEX corpus: although this \emph{search} for variability used lightcurves with 30-second resolution, the GALEX data (and the gPhoton2 software) support analysis at much higher time resolutions, down to at least 1 second, limited only by counting errors. Additionally, we believe that there are more variable sources to be found within the GALEX corpus---e.g., by searching in shorter visits or using different techniques.

New astronomical survey missions with time-domain capabilities in ultraviolet wavelengths are in the works: notably ULTRASAT \citep{ben2022scientific} and SPARCS \citep{shkolnik2016monitoring}. GFCAT suggests good candidates for targeted followup by these and other projects. However, we are not aware of any planned missions currently scoped to provide the sub-minute cadence necessary to interrogate phenomena like fine-scale morphology of ultraviolet flares, so we expect GALEX data to remain a uniquely valuable resource for many years.

\section{Acknowledgements}
The first author dedicates GFCAT \emph{in memoriam} to the lead software developer of the original GALEX direct-imaging data pipeline, Tim Conrow (1958-2017); Tim's software, mentorship, trust, and support made this work possible. We also thank Clara Brasseur, Evgenya Shkolnik, James Jackman, R.O. Parke Loyd, Eric Feigelson, Wade Roemer, and Rebekah Albach for helpful discussions over many years. We thank Michael Tucker for feedback provided during journal peer review. This research was supported by NASA grants 80NSSC18K0084 and 80NSSC21K1421. This research has made use of the SIMBAD database, operated at CDS, Strasbourg, France. This work has made use of data from the European Space Agency (ESA) mission {\it Gaia} (\url{https://www.cosmos.esa.int/gaia}), processed by the {\it Gaia} Data Processing and Analysis Consortium (DPAC,
\url{https://www.cosmos.esa.int/web/gaia/dpac/consortium}). Funding for the DPAC
has been provided by national institutions, in particular the institutions
participating in the {\it Gaia} Multilateral Agreement. All the {\it GALEX} data used in this paper can be found in MAST\dataset[10.17909/8d57-1698]{https://dx.doi.org/10.17909/8d57-1698}.

\software{Astropy \citep{astropy:2013, astropy:2018, astropy:2022},
          gPhoton1 \& 2 \citep{million2016gphoton, stclair_2022_gPhoton2},
          DAOphot \citep{stetson1987daophot},
          } 

\appendix

\section{Common categories of artifacts}
\label{artifacts_appendix}

A number of instrument artifacts and other non-astrophysical phenomena generate variable signals in the lightcurves. These are extremely common in the data---far more common than astrophysical variability---so eliminating them from consideration is, and was, the most difficult part of searching for variability in the GALEX corpus. Absent these artifacts, simple statistical cuts on the data would immediately yield reliable results, but as it is, no simple cut was possible. The characteristics of some of these artifacts in visit-level integrations have been previously described, but the intra-visit time-series behavior of these artifacts has scarcely been discussed at all in the published literature \citep{morrissey2007calibration, million2016gphoton, de2018searching} A few of them have never been described \emph{at all}, as far as we are aware, either because they were unknown or of little importance in full-depth photometric analysis. This appendix is the most comprehensive description of GALEX artifacts to date, and should provide a useful reference for anyone attempting to leverage the GALEX time domain.

Many of these artifacts can be effectively screened \emph{to some extent} by applying simple heuristics to the lightcurves---e.g. by looking for dramatic impulsive outliers, periodic data spikes, or clusters of similarly varying sources. However, we were not able to find simple algorithmic heuristics that eliminated \emph{all} such artifacts while not simultaneously eliminating many instances of clear astrophysical variability. Fortunately, almost all of these artifacts are prominently visible in properly-stretched images of either the full field of view or the target\footnote{The IRAF-created ZScale stretch, as implemented by Astropy, works quite well.}. Unfortunately, to make use of this fact, we had to manually sort through tens of thousands of false positives that still remained after our algorithmic screening processes eliminated the vast majority of our hundreds of millions of candidate source-visits. It may be possible to do better with machine learning analysis of images, especially now that we have described many categories of both true and false variability---and provided a straightforward means to produce a labeled training set for classification algorithms.

Also note that, although we are confident in our artifact screening, as \cite{million2016gphoton} emphasized, it is good practice, when making scientific use of \emph{any} imaging data, to look directly at the pictures before trusting information derived from them. The QA images included in our data supplement are intended to facilitate this process.

\begin{figure}
	\includegraphics[width=\columnwidth]{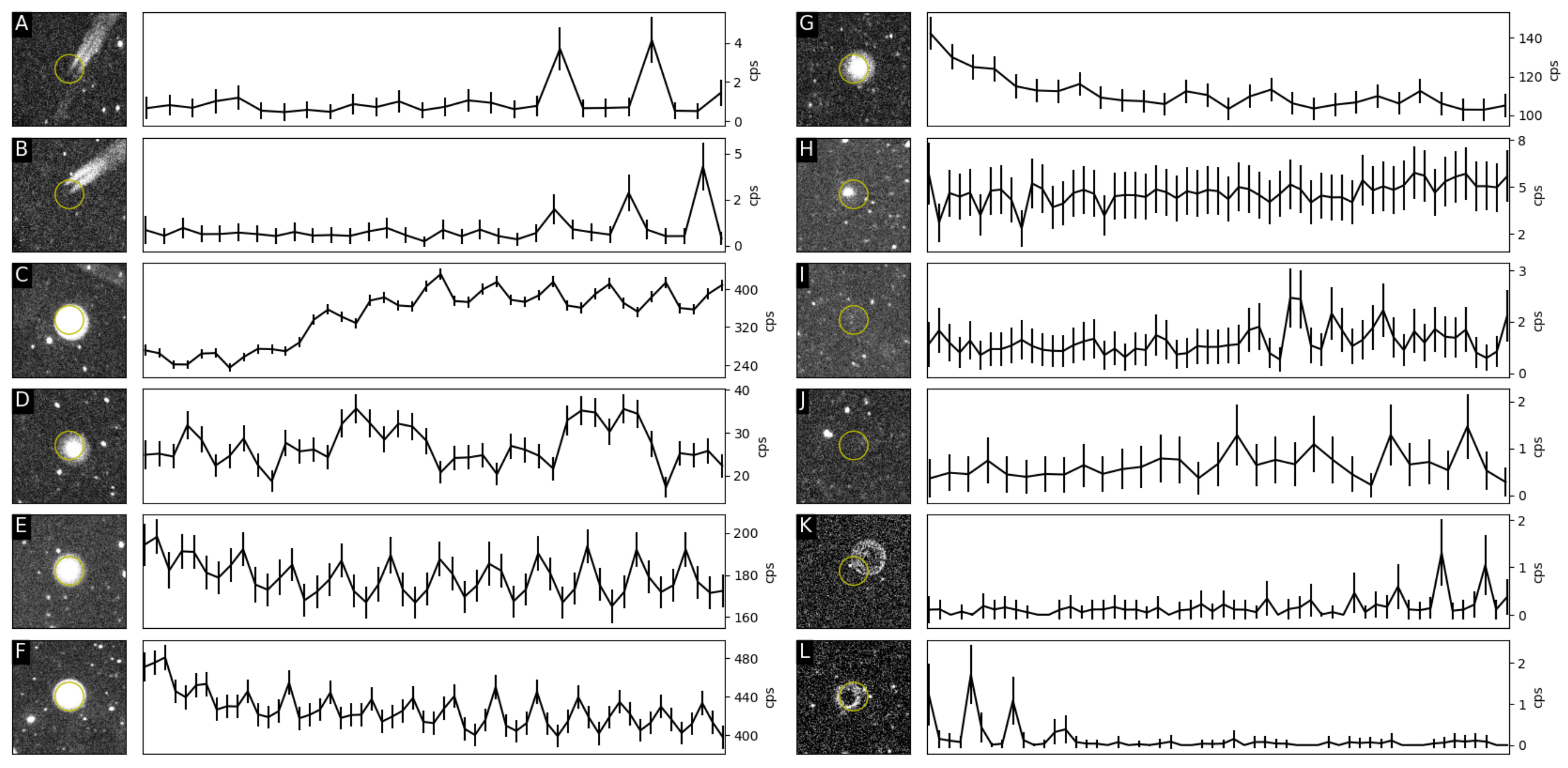}
    \caption{These are examples of ``dither-sychronous'' artifacts, which exhibit periodic signals consistent with the $\approx$120-second period of the MIS dither pattern. The time bins are 30-seconds and flux units are counts-per-second within a 17.5$''$ aperture, with 3$\sigma$ error bars. The thumbnail images are 5$'$ across; the circles in the thumbnails are just graphical indicators, not photometric apertures. Examples A and B show ``edge'' reflections. Examples C-G are all bright sources triggering non-linear response in the detector. Examples H-J show a variety of dither-like modulations on low signal or noise. Examples K and L are examples of unmasked hotspots in the FUV detector.}
    \label{fig:dither_synchronous}
\end{figure}

\begin{figure}
	\includegraphics[width=\columnwidth]{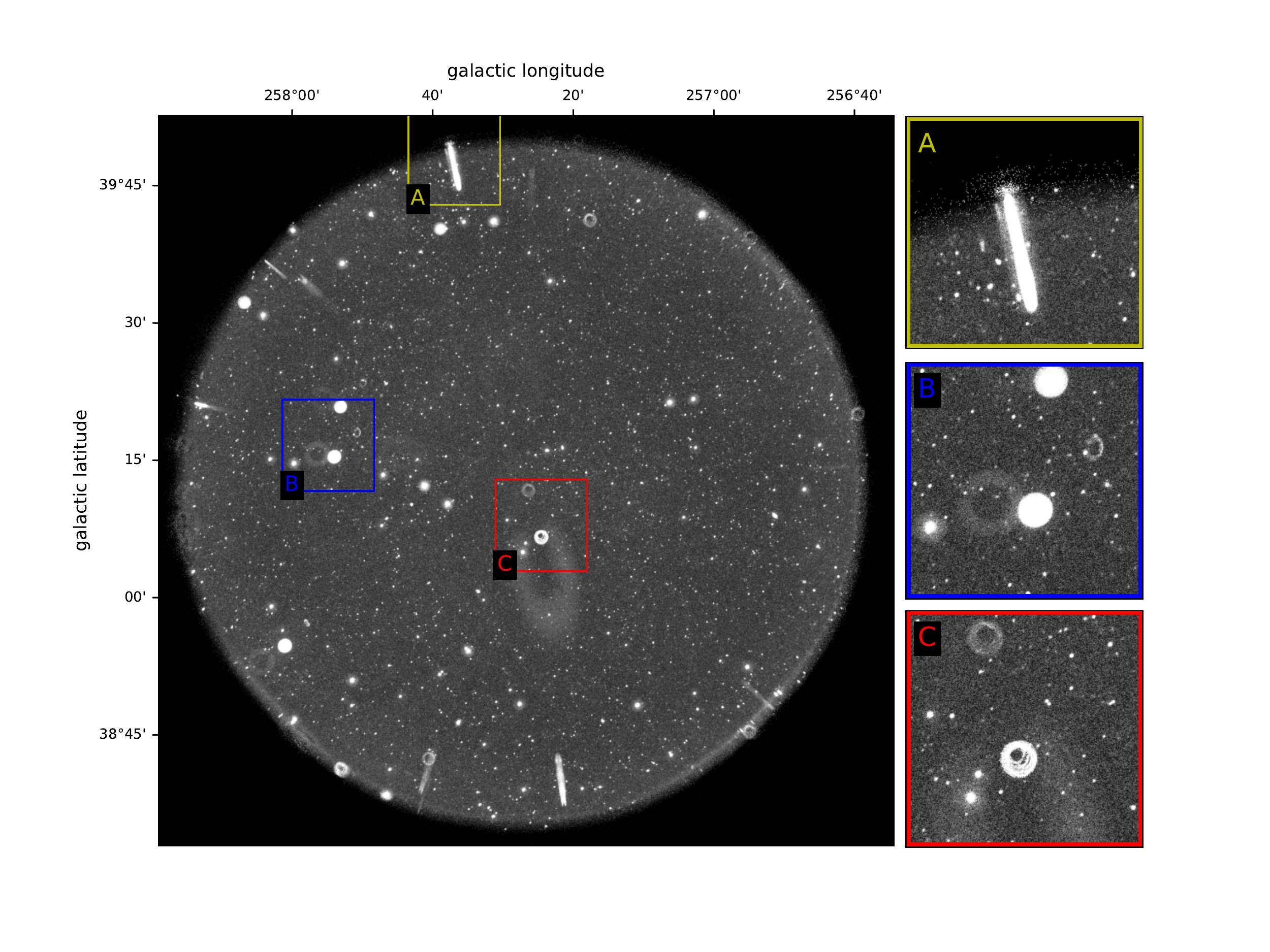}
    \caption{This is the full-depth NUV image from GALEX eclipse 1516. It contains clear examples of several common categories of artifacts. Box A contains an ``edge'' artifact that is caused by the glint of a bright star just outside of the detector field of view. Box B contains a diffuse "halo" artifact caused by the light of a relatively bright star reflecting within the telescope optical chain. Box C contains two active hotspots, characterized by a bright spiral pattern (tracing out the inverse path of the boresight dither). This region also contains part of a diffuse artifact that is an extension of the edge artifact identified in A; such artifacts sometimes extend completely across the detector.}
    \label{fig:hotspot_edge_halo}
\end{figure}

\subsection{Dither-synchronous artifacts}
In the Medium Imaging Survey (MIS) mode of observation, to which this investigation was limited, the spacecraft observed a single region of sky for 20-30 minutes. In order to avoid detector gain sag or burn in issues, the spacecraft was commanded to continuously move the boresight in a tight spiral pattern---expanding from 0.5 to 1.5 arcminutes diameter---over the course of an observation, with each loop of the spiral taking 120 seconds to complete. This spiral pattern is referred to as the ``dither pattern.'' Any signals in the GALEX data which cycle with a period of 120 seconds are therefore suspect and, indeed, many categories of artifacts can be recognized in the lightcurves by this behavior.

\subsubsection{Edge reflections}
The light from a bright star just outside the field of view of the GALEX detector can reflect in through the edge of the optics, causing a streak in the field of view. Although this is a literal ``edge'' effect, the scattered light from a star can sometimes extend across the entire field of view. These artifacts are somewhat similar to the ``Dragon's Breath'' artifacts in Hubble ACS/WFC. \citep{fowler2017analysis} As the GALEX boresight dithers over a visit, the bright star moves nearer and farther from the detector edge, causing the streak to change in intensity and extent. This means that edge reflections can generate dither-synchronous pulsations in the lightcurves of sources they cover. They can sometimes also convincingly imitate the lightcurve morphologies of stellar flares. Examples of dither-synchronous effects appear in A and B of Figure \ref{fig:dither_synchronous}. Several of these artifacts also appear in the full-frame image in Figure \ref{fig:hotspot_edge_halo}, including one that extends well into the center of the field of view.

\subsubsection{Non-linearities}
The high count rates caused by bright sources can sometimes exceed the capacity of the MCP to replenish electrons. When this happens, measured count rates cease to vary linearly with physical flux. An analysis of how local count rate suppresses measured fluxes \emph{on average} appears in \cite{morrissey2007calibration}, which gives $10\%$ rolloff values for the limit of the linear response domain---specifically, 109 cps in FUV and 311 cps in NUV. However, the edge of the linear domain actually varies by detector position (it is generally lower near the detector edges). We have found that effects can start at much lower measured count rates, even down to 100 cps in NUV. Nonlinear detector behaviors sometimes manifest as simple suppression of flux or decay, but can also generate extremely complex lightcurve morphologies, including dither-synchronous pulsations in addition to both upward and downward trends in flux. Examples C-G in Figure \ref{fig:dither_synchronous} show a variety of examples. \cite{de2018searching} also contains some additional analysis of this phenomenon.

\subsubsection{Noise modulation}
\label{modulation_of_noise}
Sometimes very faint sources or patches of apparent background also exhibit dither-synchronous variability. H-J of Figure \ref{fig:dither_synchronous} shows a few examples. This may be a noise modulation effect: as the sampled region traverses the detector flat field, variations in detector response create changes in measured count rate. This may be indicative of residual errors in the flat (previously noted as a possibility in \citep{million2016gphoton}). However, the cause is not certain.

\subsubsection{Hotspots}
Hotspots are regions of the detector that are consistently illuminated. Their precise etiology is unknown, but they are probably regions of very high response in the detector or flaws in the detector electronics that induce false signal. In GALEX image products, hotspots present as bright streaks that trace the \emph{inverse} path of the detector boresight over the course of the visit (because they are fixed in detector coordinates). In the MIS-like visits considered in this study, these traces are spiral patterns which, most of the time, create donut-like shapes. A hotspot that traverses a photometric aperture generates dramatic dither-synchronous pulsations in the resulting lightcurve. Almost all GALEX hotspots are covered by the mission-produced hotspot mask; affected sources or time-bins are therefore easy to flag and eliminate. We have discovered, however, that there are two active hotspots in the FUV band that are not covered by the mission-produced hotspot mask. Examples of these FUV hotspots appear in K and L of Figure \ref{fig:dither_synchronous}. NUV hotspots (which are all covered by the mask and therefore easily screened) appear in C of Figure \ref{fig:hotspot_edge_halo}. Spiked waveforms due to the unmasked FUV hotspots, like those that appear in the last part of K and the first part of L of \ref{fig:hotspot_edge_halo}, often occur across an entire visit, but the automated screening step described above that was intended to reject edge reflections also typically rejected this worst case.

\subsection{Post-CSP ghosts}
\begin{figure}
	\includegraphics[width=\columnwidth]{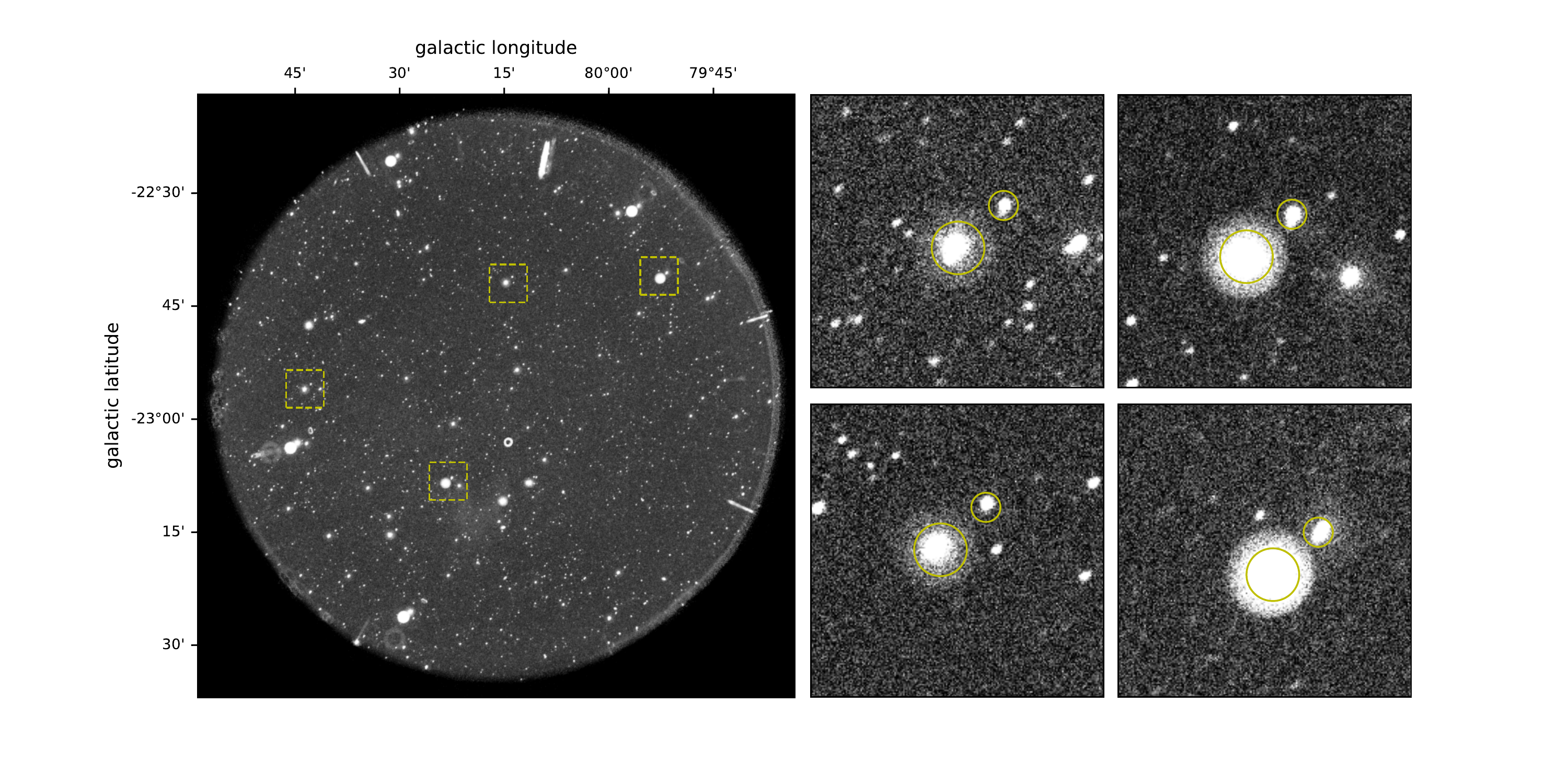}
    \caption{An example of the ``ghost'' sources that appear in images taken after the Coarse Sun Point (CSP) spacecraft anomaly. The duplicated sources arising from miscalibrated photon events are at a consistent offset \emph{within each visit} from all sources within the field, but are only visible / significant near bright sources. Note that even the bright edge reflections have ghosts. The offset is along the detector y-direction and so its vector in the sky-projected images is a function of spacecraft roll angle.}
    \label{fig:post_csp_ghosts}
\end{figure}

Following recovery from the so-called ``Coarse Sun Point'' (CSP) spacecraft anomaly during GALEX eclipse 37423, the NUV detector performance changed significantly. Due to errors in onboard electronics, the positions of a significant fraction of incident photons could no longer be accurately assigned to detector---and therefore sky---coordinates. Subsequent empirical corrections mitigated the issue, but a small fraction---less than a few percent---of incident photons were still assigned positions offset along the detector y-axis from where they should be in the image. The effect is to create, for every star in the sky, a faint ``ghost'' image many arcseconds away. When viewing a full-frame image from this part of the mission, the effect is somewhat like seeing double---because you are. See Figure \ref{fig:post_csp_ghosts} for an example.

For most photometric measurements, this matters very little. Ghosts of dim sources are generally not even detectable, because the contribution of ``misplaced'' flux is small compared to the photometric error. Ghosts of bright stars, however, can easily be mistaken for real sources, and---even more troubling for variability search---a ghost source shares any bad behaviors of its host. The most deleterious manifestation of this issue is that, if a host is very bright, its ghost, although much dimmer, will exhibit  the chaotic, non-linear behaviors typically associated with bright sources (described above and in \citet{de2018searching}). This means that, following the CSP, a simple cut on source brightness is a much less useful heuristic, and \emph{proximity} to bright stars needs to be assessed as well.

\subsection{Halos}
A ``halo'' artifact often appears near particularly bright stars, offset toward the edge of the detector at a distance that appears to vary with the distance of the star distance from the center of the field of view. These are highlighted in B of Figure \ref{fig:hotspot_edge_halo}. These artifacts sometimes induce variation in the lightcurves of sources they touch, perhaps due to slight changes in the position of the halo during the visit. The halo artifacts are believed to be caused by internal reflections in the optical chain, particularly the dichroic mirror---see a brief mention of it in \citet{morrissey2007calibration}---so they could also be called ``dichroic artifacts.'' Careful examination of the brightest examples of these artifacts reveals three radial lines evenly spaced at 60-degree offsets. These are probably the shadows of the mirror supports inside the telescope.

\subsection{FUV shadow}
\label{fuv_shadow}
\begin{figure}
	\includegraphics[width=\columnwidth]{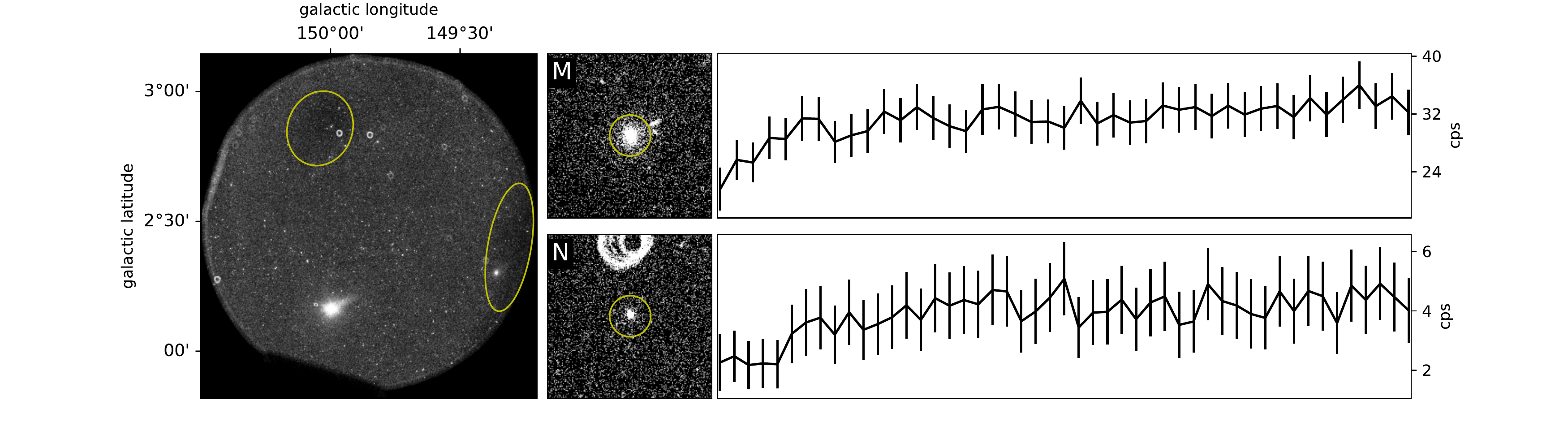}
    \caption{Highlighting two regions of the FUV detector associated with a ``ramp up'' behavior in lightcurves. The lightcurves for two sources in this visit are provided as examples of the phenomenon. The thumbnail images are 5$'$ across; the circles in the thumbnails are just graphical indicators, not photometric apertures.}
    \label{fig:fuv_shadow}
\end{figure}

We found that two regions of the FUV detector were consistently associated with ``ramp up'' behavior in sources: a phenomenon in which flux increases for the first few minutes of the visit before stabilizing. Examples of lightcurves that exhibit this behavior are given in the right panels of Figure \ref{fig:fuv_shadow}. These two regions of the detector are visible in some visits as areas of relatively dimmer background than the rest of the field of view, which supports the hypothesis that fluxes in these regions are systematically depressed during the first few minutes of the integration. We have termed these regions ``FUV shadows'' and circled them in the left panel of Figure \ref{fig:fuv_shadow} for general reference. The problem \emph{seems} to occur more frequently in visits between GALEX eclipses ca. 12889 and 13856; we don't know why. The effect might be correctable in future work, but it may be wise---if very reliable measurements are desired---to simply trim the first few minutes from FUV observations before integrating.

Note that these regions are \emph{not} the same as the region covered by the ``FUV notch'' visible in mission-produced data from the FUV detector. The notch was implemented as part of the hotspot mask to cover a low-quality region of the detector. It does not appear in our images because gPhoton2 does not apply the hotspot mask at the image level, but instead implements it as a backplane.

\subsection{Satellites}
Fast-moving satellites appear as streaks or swoops across the entire field of view. They are relatively easy to filter out. First of all, DAOPHOT does not register them as sources, so they only affect any sources they transit. The lightcurves of sources they transit often exhibit dramatic jumps in brightness \emph{for exactly one bin}---two at most---which is behavior that our screening algorithms could and did detect. When the effect is not dramatic enough to be automatically screened, it can still be seen easily in the QA images and filtered manually.

\begin{figure}
	\includegraphics[width=\columnwidth]{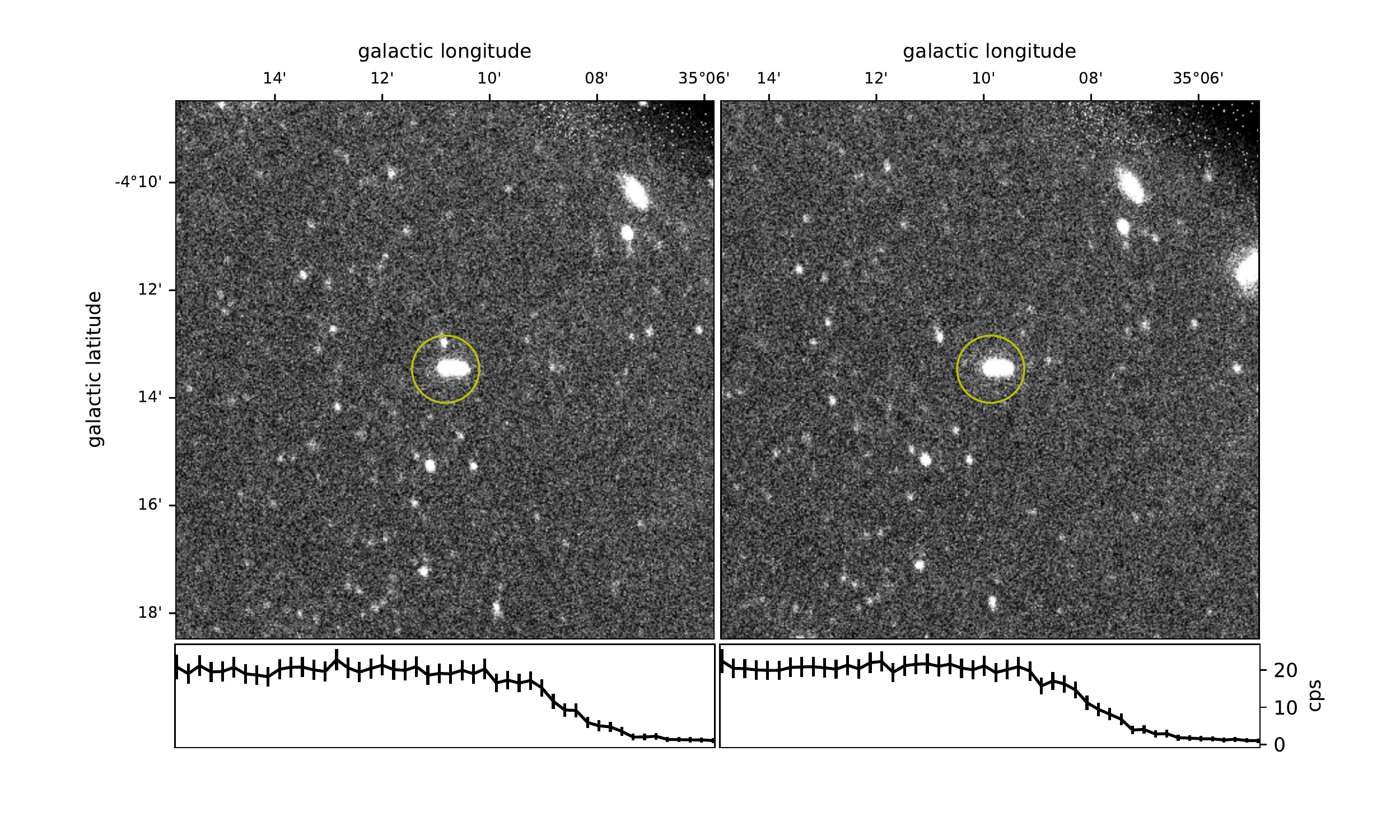}
    \caption{As an example of slow-moving artificial satellites in the GALEX data, this is the same object observed in back-to-back GALEX eclipses 7958 and 7959. The decaying lightcurve, caused by the source moving across and out of the photometric aperture over the course of the visit, combined with the elongated source, is diagnostic of such objects.}
    \label{fig:slow_moving_objects}
\end{figure}

Slow-moving objects are more difficult. These might drift only a few pixels in the image---maybe less than $10''$---over the course of an entire observation. Also, DAOPHOT detects them as sources, which means we produced lightcurves for them. The lightcurve of an object like this always exhbits variation, because the object changes position with respect to any photometric aperture over the course of a visit. In full-frame images, they look like slightly misshapen point sources---which, unfortunately, is not particularly alarming, because many GALEX sources are galaxies with small but not point-like angular extents or stars that become slightly misshapen near the detector edges. Slow-moving objects are somewhat more easily identified in QA \emph{animations}. However, we were able to determine that they have a characteristic lightcurve morphology: a sigmoid-like decay (shown in Figure \ref{fig:slow_moving_objects}). The combination of this lightcurve morphology with a slightly elongated point source is diagnostic.

We have simply assumed that these are artificial satellites. The GALEX detectors are particularly good for ``space sensing'' applications (see \citet{siegmund2006advanced} for a discussion). However, it is possible that some of them are near-Earth objects with non-terrestrial origins. An exploration of this possibility was outside the scope of this work.

\subsection{The special problem of flux decay}
The category of artifact demonstrated by examples E, F, and G in Figure \ref{fig:dither_synchronous}, in which the rate of flux of moderately bright sources drops in an approximately exponential decay over the course of the observation, was a new, unexpected, and rather annoying discovery in the course of this project. The cause of this behavior is uncertain. Local detector gain sag---depletion of the portion of the MCP in the immediate vicinity of a star that overwhelms its ability to replenish electrons---is the most plausible hypothesis. However, the spacecraft dither was supposed to mitigate this, it occurs well below flux thresholds previously identified as problematic, and the effect is more subtle and less chaotic than the ``non-linearity'' behaviors described in \citet{de2018searching}. Absent certainty about the mechanism, we have decided to call it simply ``flux decay.''

\begin{figure}
	\includegraphics[width=\columnwidth]{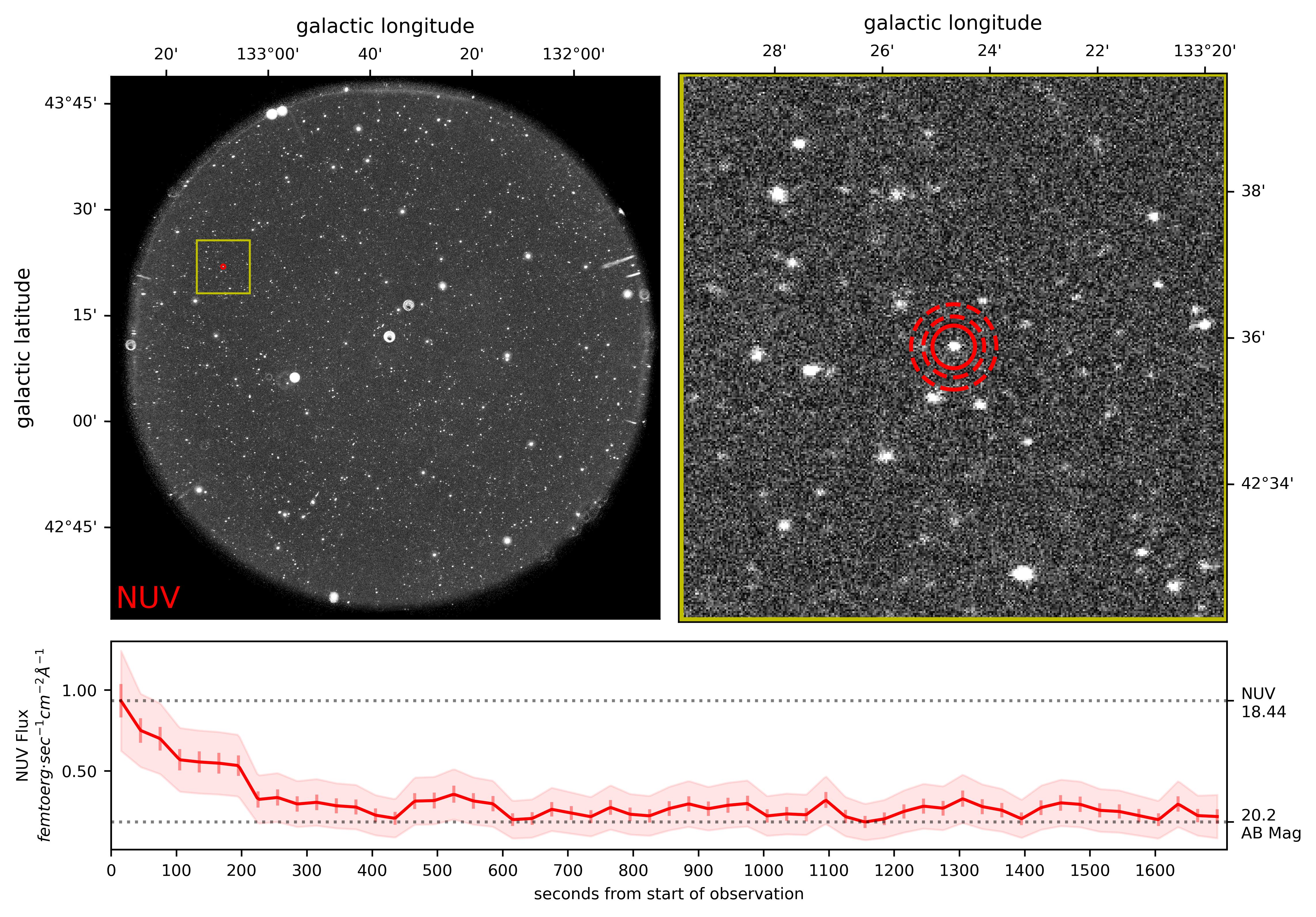}
    \caption{GFCAT J085234.4+432202.3, observed in GALEX eclipse 35852, presents with a lightcurve morphology that is plausibly astrophysical but also mimics known categories of non-astrophysical artifacts. Refer to the capture of Figure \ref{fig:gfcat_07500_0205} for a description of the plot layout.}
    \label{fig:gfcat_35852_1273}
\end{figure}

The trouble it presents to intra-visit variability search is that the artifact appears to impact some---but not all!---sources previously considered well within GALEX's safe brightness thresholds, and often creates lightcurve morphologies that might be \emph{perfectly plausible} to associate with astrophysical variability (e.g., the exponential decay phase of a large flare). For example, the lightcurve of the object shown in Figure \ref{fig:gfcat_35852_1273}, observed in GALEX eclipse 35852, has exactly the same morphology as lightcurves affected by by gain sag. The fact that its brightness is only 18.44 mag makes it seem implausible that it is due to a detector gain sag issue---it is more likely the tail end of some other trend with a stellar cause. It might be the tail end of a flare, the entry phase of a stellar eclipse, or part of a $\delta$ Scuti pulsation. This source has no good matches in SIMBAD or ViZieR, so we can only guess.

\begin{figure}
	\includegraphics[width=\columnwidth]{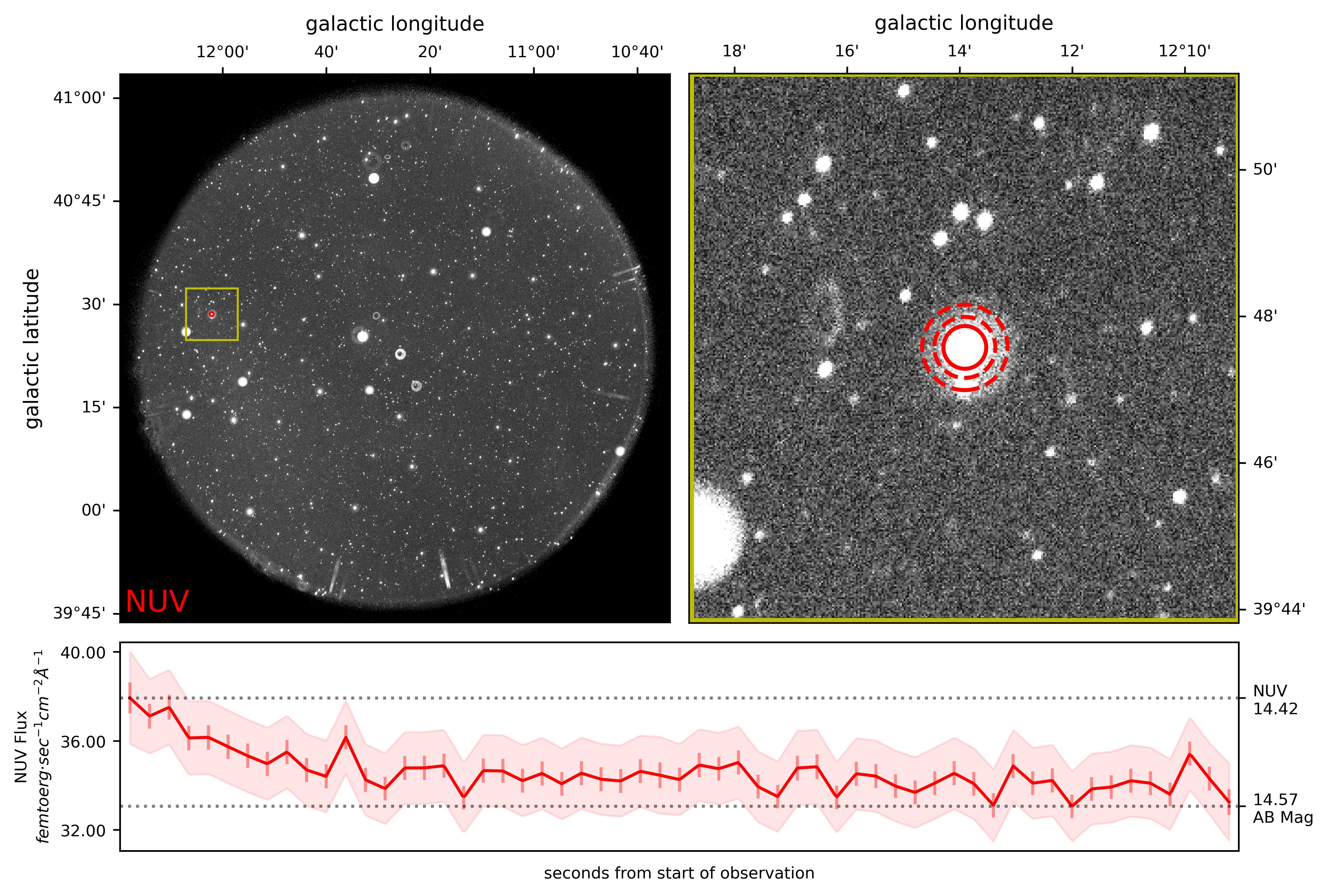}
    \caption{This target (TYC 2801-218-1) observed in GALEX eclipse number 34581 demonstrates ``decay-like'' behavior that might be the decay phase of a flare, but was eliminated from GFCAT as likely due to a detector artifact. Refer to the capture of Figure \ref{fig:gfcat_07500_0205} for a description of the plot layout.}
    \label{fig:gfcat_34581_1233}
\end{figure}

\begin{figure}
	\includegraphics[width=\columnwidth]{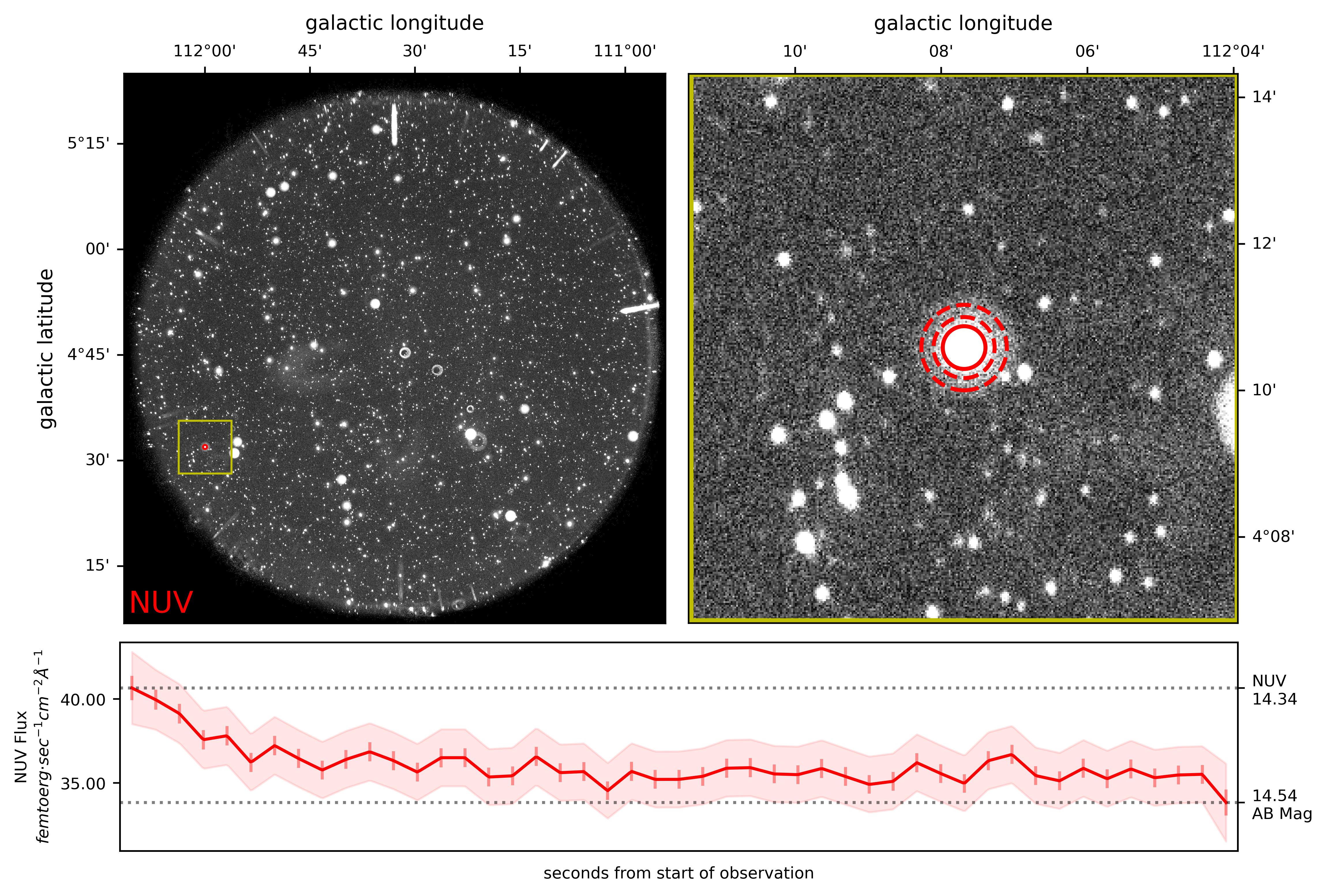}
    \caption{This target observed in GALEX eclipse number 35709 demonstrates ``decay-like'' behavior that might be the decay phase of a flare, but we eliminated from GFCAT as likely due to a detector artifact.}
    \label{fig:gfcat_35709_1265}
\end{figure}

In some cases---especially because these are relatively bright sources---it is possible to lean a little bit on prior identifications. For example, see Figure \ref{fig:gfcat_34581_1233}, depicting an object observed in GALEX eclipse 26808. This object was previously observed by Tycho (TYC 2801-218-1) and has a SIMBAD object type classification of simply "Star." It seems unlikely (although certainly not impossible)  that if an object this bright produced flares large enough for this to be a flare decay phase, it would not already have been identified as variable, so we eliminated the visit from the sample. But the object in Figure \ref{fig:gfcat_35709_1265}, observed only once in GALEX eclipse 35709 is trickier. It has the same flux decay shape---going from 14.34 to 14.54 AB mag---and is listed in SIMBAD as a known binary star system. Binary star systems are sometimes very active, so it is not out of the question that, given no additional context, this morphology might show the decay phase of a flare or the start of a stellar eclipse. Nonetheless, we chose to exclude it out of caution.

\begin{figure}
	\includegraphics[width=\columnwidth]{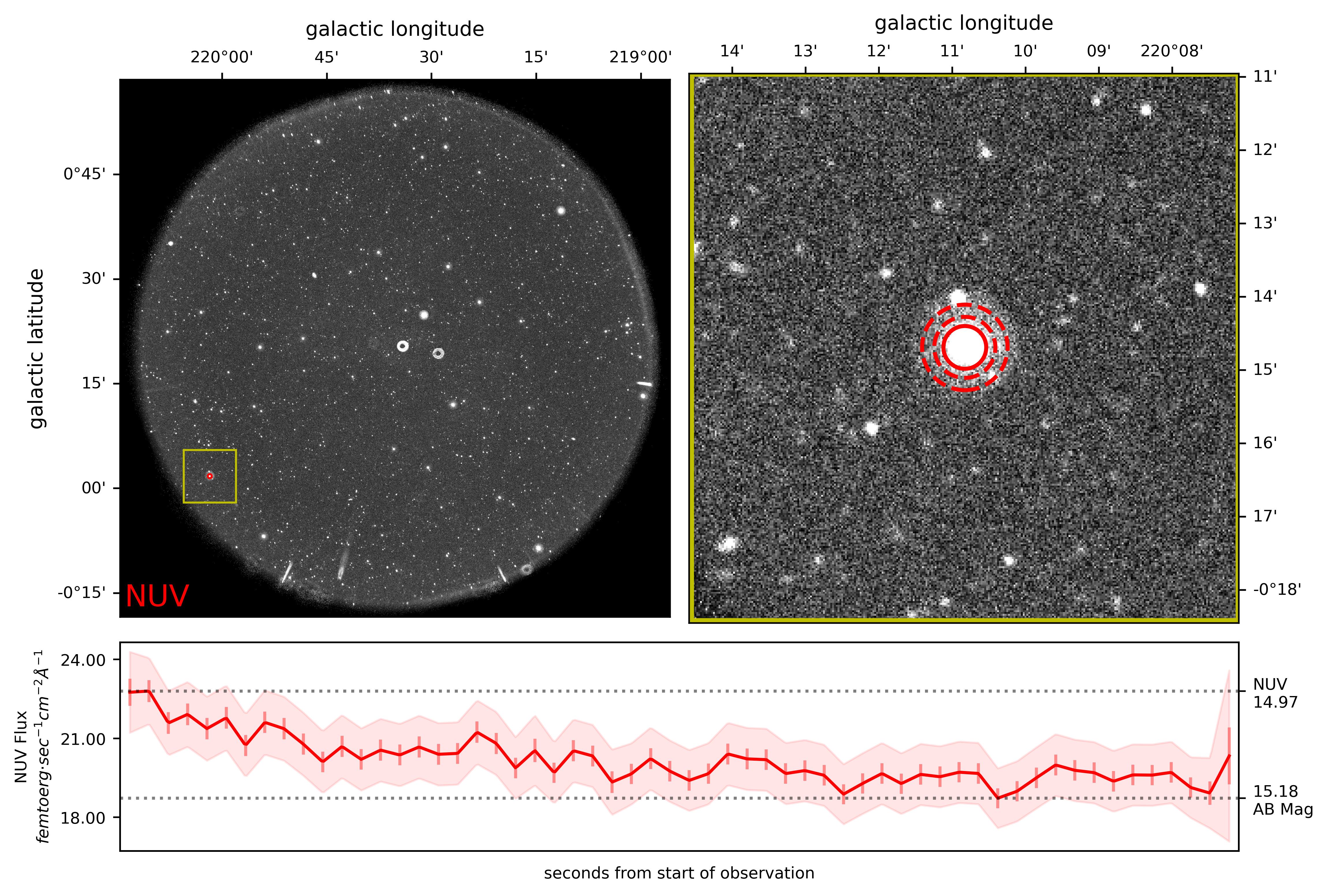}
    \caption{GFCAT J144007.1-000143.8 is the well-studied $\delta$ Scuti variable IP Virgo. Although it presents with a light curve morphology that is plausibly either astrophysical or non-astrophysical in origin. The prior knowledge that this is a known variable increases our confidence in its inclusion in GFCAT. Refer to the capture of Figure \ref{fig:gfcat_07500_0205} for a description of the plot layout.}
    \label{fig:gfcat_21232_0812}
\end{figure}

And then there is GFCAT J144007.1-000143.8 (shown in Figure \ref{fig:gfcat_21232_0812}), observed in GALEX eclipse 21232. It has all of the hallmarks of flux decay, including the right morphology and being in the approximately-correct range of brightness. It even appears somewhat near the edge of the detector. And yet this source is the well-studied $\delta$ Scuti star IP Virgo! Given its stellar type, we should \emph{certaintly} not be at all surprised to find dramatic fast variations of precisely the type shown here. We chose to leave it in GFCAT with the QA flag set. But we cannot determine with certainty whether or not it is legitimately variable within this GALEX visit.

We did undertake some systematic exploration of this flux decay effect to try to establish relationships between source brightness, source position, proximity to the detector edge, etc. Unfortunately, we have not been able to find any reliable correlation or cause so far. It does \emph{appear} to be more common ``near-ish'' the detector edges, but there is also more observing area at larger radii---in other words, there are simply more stars close to the edge of the detector than not---so this might simply be a sampling bias. We chose GFCAT's $\approx170$ counts-per-second ($\approx14.4$ AB mag) soft cutoff for source brightness (in NUV) precisely because it seemed---from inspection of tens of thousands of lightcurves---to be about the ballpark brightness at which this decay behavior became common.

\subsubsection{Hints of consequences for GALEX calibration generally}
\citet{camarota2014white} and \citet{wall2019galex} have suggested that the GALEX absolute calibration might be systematically offset: too dim by a few percent, and worse for brighter stars. The flux decay behavior that we observed supports concerns that reported GALEX photometric measurements for some sources may be too low. To wit: the primary GALEX calibration standard star---which was the basis for the photometric baseline and, at least initially, the detector flat fields---was the white dwarf LDS749B, with GALEX brightness of about 14.76 AB mag in NUV and 15.6 AB mag in FUV. This translates to about 145 counts-per-second in NUV, which is well inside the brightness range in which we observed the flux decay behavior. The majority of calibration observations were taken in a special calibration mode (called ``Calibration Imaging'' or ``CAI''). In this mode, the spacecraft boresight shifted every $\approx120$ seconds over the course of a GALEX eclipse in such a way that this standard star would land on (or sample from) a different part of the detector. These calibration observations were, generally, only about 2 minutes long, but an overwhelming fraction of the total data corpus consists of observations over 20 minutes long. Although the decay behavior is \emph{present} from the very beginning of the lightcurve, it is negligible---maybe even imperceptible---until a few minutes into the visit. (120 seconds corresponds to only the first 4 bins of the 30-second resolution lightcurves shown throughout this paper.) Integrating the flux of a source affected by this decay behavior over only the first two minutes of a visit yields a brighter measurement (of, effectively, mean flux) than integrating over a full GALEX visit. This could translate to systematically-offset flux measurements through a variety of mechanisms. For instance, it could could bias calibration or introduce inaccuracies into the flat field. It would also tend to cause flux measurements of affected stars integrated over longer visits to be depressed by at least a few percent; this ``local dead time'' effect was noted as a possibility by \citet{camarota2014white}, and seems plausible. Although by \citet{camarota2014white} and \citet{wall2019galex} achieved seemingly good corrections by comparing GALEX photometry to models, our experience suggests that there may be a dependence on observation exposure time and detector region as well. Further investigation of the nature, extent, and cause of the flux decay behavior is warranted.

We think it's worth reemphasizing that the nonlinearity analysis in \citet{morrissey2007calibration} was performed specifically on white dwarf standard stars \emph{mostly if not entirely observed in CAI mode}, which was non-representative of the normal scientific observation modes. It therefore plausibly \emph{overestimates} the threshold at which significant nonlinearity effects kick in during longer, MIS-like observations. Many people---us included---have historically used the ``$10\%$ rolloff'' value reported in this paper (109 cps in FUV and 311 cps, or 13.8 AB mag in NUV)---as a ``safe'' threshold of brightness for GALEX photometry. (One might also be tempted to simply add $10\%$ flux back into the measurement, or to propagate $10\%$ more error. Neither of these are good ideas. As, e.g. \citet{de2018searching} showed, and as we have seen, behaviors in the nonlinear regime are often far more complicated than simple response rolloff.) However, the threshold may in fact be much dimmer, especially in the case of full-depth GALEX observations. \citet{wall2019galex} found that a correction may be needed for stellar brightness as low as 15.95 AB mag in FUV and 16.95 AB mag in FUV, corresponding to countrates of only about 17.9 and 14.1 cps. We still believe that the GALEX archival data are generally reliable and of very high quality, but they should---as with all data---be interpreted in the context of awareness that our instruments are as mysterious and imperfect as we are.

\clearpage
\bibliography{gfcat}{}
\bibliographystyle{aasjournal}



\end{document}